\documentclass[journal,10pt]{IEEEtran}
\usepackage[ruled,vlined]{algorithm2e}
\usepackage{dblfloatfix}

\ifCLASSINFOpdf

\else

\fi
\usepackage{amsfonts,epsfig,cite,array,multirow,graphicx,amsmath,ltablex,tabularx,setspace,amssymb,multirow}
\usepackage{schemabloc,tikz,amsthm}
\usepackage{array}
\usepackage[caption=false,font=normalsize,labelfont=sf,textfont=sf]{subfig}
\usepackage[small]{caption}
\usepackage[numbers,sort&compress]{natbib}
\usetikzlibrary{circuits, arrows}
\usepackage{verbatim}
\usepackage{graphicx}
\usepackage{epsfig}
\usepackage{epstopdf}
\usepackage{graphicx}
\usepackage{amsmath}
\usepackage{enumitem}
\usepackage{amssymb}
\usepackage{graphics}
\usepackage{cite}
\usepackage{booktabs}
\usepackage{amsfonts}
\usepackage{textcomp}
\usepackage{multicol}
\usepackage{multirow}
\usepackage{color}
\usepackage{fixltx2e}
\usepackage{latexsym}
\usepackage{afterpage}
\usepackage{a4wide}
\usepackage{multirow}
\usepackage{float}
\usepackage{url}
\newtheorem{example}{Example}
\usepackage{hyperref}
\usepackage{subcaption}
\usepackage{schemabloc,tikz,amsthm}
\usepackage{subfig}
\usepackage{tabularx}
\usepackage[numbers,sort&compress]{natbib}
\usetikzlibrary{circuits, arrows}
\usepackage{lipsum}
\usepackage[top=.7in, left=.5in, right=.5in, bottom=.7in]{geometry}
\usepackage{xcolor}% or package color
\usepackage{tikz}
\usetikzlibrary{plotmarks}
\usepackage{pgfplots}
\usetikzlibrary{spy}
\usepackage{collectbox}
\usepgfplotslibrary{groupplots}
\usetikzlibrary{spy}
\usepackage{pgfplots}
\pgfplotsset{compat=1.18} % Ensure compatibility with your PGFPlots version
\usepackage{soul,color}
\usepackage[none]{hyphenat}

\tikzset{new spy style/.style={spy scope={%
			magnification=2.8,
			size=1.25cm,
			connect spies,
			every spy on node/.style={
				rectangle,
				draw,
			},
			every spy in node/.style={
				draw,
				rectangle,
			}
		}
	}
}

\pgfplotscreateplotcyclelist{mycolorlist}{%
	black,densely dashed,every mark/.append style={fill=black!80!black},mark=o\\%  for ideal CE
	brown,every mark/.append style={fill=brown!80!black},mark=otimes\\%
	red,every mark/.append style={fill=red!80!black},mark=otimes\\%
	blue,every mark/.append style={fill=blue!80!black},mark=otimes\\%
	magenta,every mark/.append style={fill=blue!80!black},mark=otimes\\%
	red,densely dashed,every mark/.append style={fill=black!80!black},mark=o\\%  for ideal CE
	blue,densely dashed,every mark/.append style={fill=black!80!black},mark=o\\%  for ideal CE
	brown,densely dashed,every mark/.append style={fill=black!80!black},mark=o\\%  for ideal CE
}

\makeatletter
\newcommand\footnoteref[1]{\protected@xdef\@thefnmark{\ref{#1}}\@footnotemark}
\makeatother
\begin{document}
	\tikzset{new spy style/.style={spy scope={%
			magnification=2.8,
			size=1.25cm,
			connect spies,
			every spy on node/.style={
				rectangle,
				draw,
			},
			every spy in node/.style={
				draw,
				rectangle,
			}
		}
	}
}
\SetKwRepeat{Do}{do}{while}
\newcolumntype{L}[1]{>{\raggedright\let\newline\\\arraybackslash\hspace{0pt}}m{#1}}
\newcolumntype{C}[1]{>{\centering\let\newline\\\arraybackslash\hspace{0pt}}m{#1}}
\newcolumntype{R}[1]{>{\raggedleft\let\newline\\\arraybackslash\hspace{0pt}}m{#1}}
\pgfdeclarelayer{background}
\pgfdeclarelayer{foreground}
\pgfsetlayers{background,main,foreground}
\newcommand{\oeq}{\mathrel{\text{\sqbox{$=$}}}}
\setlength{\textfloatsep}{0.1cm}
\setlength{\floatsep}{0.1cm}
\tikzstyle{int}=[draw, fill=white!20, minimum size=2em]
\tikzstyle{init} = [pin edge={to-,thin,black}]
	\title{Hybrid Rate-Splitting and Sparse Code Multiple Access (RS-SCMA): Design and Performance}
    \author{Minerva Priyadarsini,~\IEEEmembership{Student Member,~IEEE,}%
    ~Zilong Liu,~\IEEEmembership{Senior Member,~IEEE,}
        ~Kuntal Deka,~\IEEEmembership{Member,~IEEE,}%
        ~Sujit Kumar Sahoo,~\IEEEmembership{Senior Member,~IEEE,}%
        ~and Sanjeev Sharma,~\IEEEmembership{Senior Member,~IEEE}%
        \thanks{Minerva Priyadarsini and Sujit Kumar Sahoo are with the School of Electrical Sciences, Indian Institute of Technology Goa, Goa 40301, India (email: minerva183212005@iitgoa.ac.in; sujit@iitgoa.ac.in).}%
        \thanks{Zilong Liu is with  is with the School of Computer Science and Electrical
Engineering, University of Essex, Colchester CO4 3SQ, U.K. (e-mail:
zilong.liu@essex.ac.uk).}%
        \thanks{Kuntal Deka is with the Department of EEE,  Indian Institute of Technology Guwahati, Guwahati 780139, India (email: kuntaldeka@iitg.ac.in).}%        
        \thanks{Sanjeev Sharma is with the Department of Electronics Engineering,  Indian Institute of Technology (BHU), India (email: sanjeev.ece@iitbhu.ac.in).}%
}

  %   \begin{comment}
  %       \author{ Minerva Priyadarsini$^{1}$, Kuntal Deka$^{2}$, Zilong Liu$^3$, Sujit Kumar Sahoo$^{1}$, and Sanjeev Sharma$^4$\\
		% $^1$Indian Institute of Technology Goa, India,   $^2$Indian Institute of Technology Guwahati, India, $^3$University of Essex, Wivenhoe Park, Colchester, United Kingdom, $^4$Indian Institute of Technology Varanasi, India \\
  %   \end{comment}
	
		\vspace{-0.5in}

	
	\maketitle
	
	\begin{abstract}
    This paper proposes, for the first time, a hybrid multiple access framework that integrates the principles of rate-splitting (RS) and sparse code multiple access (SCMA) in an SISO downlink scenario. The proposed scheme, termed RS-SCMA, unifies the powerful interference management capability of rate-splitting multiple access (RSMA) with the near-optimal multiuser detection of  SCMA. A key feature of RS-SCMA is a tunable splitting factor $\alpha$, which governs the allocation between the generic $M$-ary modulated common messages and SCMA-encoded private messages. This enables dynamic control over the fundamental trade-off between system sum-rate, bit error rate (BER), and the overloading factor. We develop novel transmitter and receiver architectures based on soft successive interference cancellation (SIC), incorporating message passing algorithm (MPA) detection and soft-symbol reconstruction. Furthermore, a unified analytical expression for the achievable sum-rate is derived as a function of the splitting factor $\alpha$. The performance of the proposed RS-SCMA system is evaluated in terms of both BER and sum-rate. Simulation results confirm the superiority of RS-SCMA over conventional SCMA and multi-carrier RSMA, demonstrating its scalability and robustness even in the presence of channel estimation errors.
\end{abstract}
\section{Introduction}
\label{sec:introduction}

\subsection{Background and Related Works}
Emerging wireless technologies are rising to meet the escalating demands for extreme data rates, massive connectivity, ultra-reliable performance, and near-zero latency, propelling the evolution toward sixth-generation (6G) networks and beyond \cite{6g_survey, 6g_tech, 6g_acces, 6g_new2}. Among many others, a fundamental and pressing research problem is how to design highly efficient, reliable, and scalable multiple access (MA) schemes that can meet diverse requirements from tens and billions of machine-type communication devices and applications. In the context of this background, rate-splitting multiple access (RSMA) has emerged as a promising scheme for 6G. By effectively managing interference through rate splitting (RS) and successive interference cancellation (SIC), RSMA provides a unified framework for enhanced spectral efficiency, reliability, and user fairness \cite{Mao2022},\cite{rsma_primer}. %Recent studies in \cite{Mao2022,rsma_primer} show that RSMA offers a flexible and efficient approach to interference management, consistently outperforming traditional multiple access techniques in various communication scenarios.

The concept of RS was first introduced in \cite{RS_Gaussian} for Gaussian multiple access channels. In an $M$-user Gaussian channel, up to $2M-1$ independent virtual channels are created for $2M-1$ ``virtual sources" enabling the partitioning of each user's rate and giving rise to RSMA.
An extension of RSMA was presented in \cite{rs_2001}, where users' data are divided into multiple streams, termed ``virtual users", in the context of discrete memoryless channels (DMC). It was demonstrated that rate-splitting enables the achievement of any rate within the capacity region of a DMC.%, thus paving the way for contemporary RSMA research. % enhancing multi-user communication systems. It addresses challenges such as maximizing spectral efficiency and managing interference effectively.

Building on this foundational principle, further advances in RSMA have been explored. The authors of \cite{Dizdar2020} optimized RSMA by focusing on physical layer design, modulation and coding schemes, precoder design, and efficient message splitting, highlighting its adaptability to diverse requirements in modern communication networks. A link-level performance evaluation for downlink RSMA systems was carried out in \cite{Chen2020}, with a particular emphasis on error performance analysis. The work in \cite{sic_free_conf} explored SIC-free receiver designs by taking advantage of an optimized precoder for finite constellations applicable to both SIC-based and SIC-free RSMA. More practical receiver architectures were developed in \cite{rsma_receiever} for RSMA under finite constellation. A variety of low-complexity precoders and different receiver options were proposed, covering both SIC and non-SIC based designs. %In \cite{rsma_ofdm}, RSMA was integrated with orthogonal frequency division multiplexing (OFDM) to address inter-carrier interference (ICI) and inter-symbol interference (ISI). The study carried out optimal power and subcarrier allocation through sum-rate maximization, further underscoring RSMA's potential in advanced wireless systems.
\cite{mc_rsma} studied resource allocation for downlink multicarrier RSMA (MC-RSMA) by jointly optimizing the per-subcarrier power split, user–subcarrier matching, and inter-subcarrier power allocation to maximize the system sum rate. In \cite{Pereira2025}, the authors investigated subcarrier allocation for downlink MC-RSMA by jointly optimizing user matching, subcarrier assignment and stream-based power allocation to maximize the weighted sum rate (WSR). Joint power and subcarrier allocation were studied in \cite{Chen2020_mc_rsma} for a multicarrier multigroup multicast multiple-input-single-output (MISO) downlink system using RS to manage inter-group interference in overloaded scenarios. Further, RS was applied in \cite{Chen2021} for an overloaded multicarrier multigroup multicast downlink system by optimizing the transmitter design (along with subcarrier/power allocation) to improve max–min fairness and coded BER performance.

In parallel with the RSMA research, significant efforts have been devoted to code-domain non-orthogonal multiple access (CD-NOMA) schemes, with sparse code multiple access (SCMA) emerging as a prominent candidate \cite{Liu2021}. Over the past decade, SCMA has attracted widespread interest due to its ability to achieve error-rate performance close to that of a maximum-likelihood receiver with low decoding complexity \cite{SCMA1, scma_cb}. The key innovation of SCMA lies in its use of sparse codebooks and the message passing algorithm (MPA), which enables robust and efficient multi-user detection over heavily loaded massive machine-type communication networks.

SCMA provides distinct advantages over other CD-NOMA techniques, particularly in terms of the constellation shaping gain and overloading flexibility. Codebook design plays a central role in SCMA system development and is typically based on transforming a multi-dimensional mother constellation using operations such as phase rotation, permutation, and interleaving \cite{scma_cb, Zhang2016, Yu2018, Zeina2018, Chen2020b, ioT_scma}. These transformations yield user-specific codebooks that strike a balance between sparsity and minimum distance properties. For example, power-imbalanced SCMA codebooks were proposed in \cite{Li2022} for achieving enhanced codebook minimum distance properties by allocating different power levels across multiple users. A comprehensive tutorial on SCMA detection and decoding methods was provided in \cite{scma_zilong}, which systematically outlines algorithmic variants and practical considerations for implementation. %Together, these advances underscore SCMA’s potential as a robust and scalable multiple access scheme, particularly suited for overloaded, interference-limited scenarios such as satellite and industrial internet-of-things (IoT) networks as well as densely connected vehicular networks.

  \subsection{Motivation and Contributions}
  {\color{black}
%While RSMA has been extensively investigated alongside power-domain NOMA (PD-NOMA)~\cite{NOMA_survey, NOMA_survey1}, its integration with code-domain NOMA (CD-NOMA) remains an unexplored research frontier. The current literature identifies RSMA as a cornerstone of a futuristic universal multiple access (UMA) architecture~\cite{uma}, yet a concrete RSMA-CD-NOMA design is still missing. Within the CD-NOMA family, SCMA has emerged as a particularly robust scheme, due to its sparse multidimensional complex codebooks, which achieve high overloading while maintaining low-complexity detection via MPA. 
A distinctive feature of RSMA is that it generalizes and unifies spatial division multiple access (SDMA), power-domain non-orthogonal multiple access (PD-NOMA) \cite{NOMA_survey, NOMA_survey1}, and physical-layer multicasting to explore operating points that are not achievable by any of them. Building upon this observation, the concept of univeral multiple access (UMA) was first coined in \cite{uma}. Compared to the existing MA schemes (e.g., RSMA, SCMA), UMA should be able to exploit all dimensions of time, frequency, power, space (e.g., antennas and beams), and signal (e.g., messages and codes) to provide intelligence and multifunctionality for the upcoming 6G networks and beyond. In particular, it was mentioned in \cite{uma} that ``\textit{UMA should further shrink the knowledge tree of MA schemes by unifying RSMA with all other dimensions, such as code-domain MAs, and ultimately provide a unified and conceptually simple understanding of the current and future morass of MA schemes. Such UMA does not exist yet}."

Motivated by \cite{uma}, this work aims to exploit the synergies of RS and SCMA for the ``first step" of the UMA study. However, the integration of RS and SCMA is non-trivial because the SIC decoding in RSMA may fundamentally disrupt the MPA decoding for SCMA. A naive hard-decision SIC performed prior to the MPA could distort the decoding convergence, thus resulting in catastrophic error propagation and significant performance degradation. Such an architectural conflict that prevents the effective harmonization of these two paradigms will be addressed by this work. Secondly, contemporary RSMA generally relies on multi-antenna techniques to carry out precoding, but this requires channel state information at the transmitter (CSIT). By contrast, SCMA only requires statistical channel state information for optimal sparse codebook design. 

%Conventional RSMA employs SIC that relies on the reliable decoding and subtraction of a common layer. In contrast, SCMA detection is performed via the MPA, whose convergence depends on the statistical properties of the full, unaltered superimposed signal across the factor graph. 

Despite these challenges, RSMA and SCMA offer complementary strengths that address the limitations of each scheme in isolation. In single-input-single-output (SISO) or low-rank channels, RSMA has limited ability to scale the number of simultaneously served users due to the lack of spatial dimensions for private streams. Conversely, SCMA achieves high overloading through code-domain multiplexing, yet it lacks an adaptive mechanism to flexibly manage the multi-user interference faced by different users.  %mitigate the interference-driven error floors that emerge in heavily loaded regimes. 
A unified RS-SCMA design harmonizes these paradigms, allowing SCMA to benefit from the additional multiplexing dimensions required in resource-constrained settings, while utilizing the common/private structure of RSMA to decode and cancel the multi-user interference that leads to more flexible interference management.

The key contributions of this paper are summarized as follows:

\begin{itemize}
\item We propose, for the first time, a hierarchical transmission structure that superposes an $M$-QAM-modulated common stream with SCMA-encoded private streams, enabling the multi-user interference to be partially decoded and partially cancelled while permitting MPA to detect all the private messages. A tunable splitting factor $\alpha$ is introduced to allocate the message bits between these two layers, enabling dynamic control of the effective overloading factor. We derive the relationship between the splitting factor, the achievable overloading factor, and the overall system spectral efficiency.

  %  \item  We propose, for the first time, a unified RSMA architecture in the code domain. The transmitter superimposes an $M$-QAM-modulated common message with SCMA-encoded private messages. A tunable splitting factor $\alpha$ allocates information between these two layers, enabling dynamic control of the effective overloading factor.

    %\item We derive the relationship between the splitting factor, the achievable overloading factor, and the overall system spectral efficiency. The proposed RS-SCMA architecture is benchmarked against conventional RSMA and SCMA systems.

    \item  We design two novel receivers that integrate soft SIC into the SCMA detection framework: 1) A low-complexity receiver (i.e., Rx-1) for uncoded RS-SCMA systems that uses the log-likelihood ratios (LLRs) from the QAM demodulator to perform soft SIC before the iterative MPA detection; 2) An enhanced receiver (i.e., Rx-2) for coded RS-SCMA systems, which uses refined soft bits from the channel decoder to improve SIC accuracy, creating a powerful decoder-assisted feedback loop. We analyze and compare the complexity and coded performance of both receivers.

    \item  We derive a unified analytical expression for the achievable sum-rate of RS-SCMA as a function of $\alpha$, for both perfect and imperfect SIC. This framework reveals the fundamental trade-offs among spectral efficiency, BER, and system overloading factor. The analysis is further extended to incorporate the impact of imperfect channel state information at the receiver (CSIR) through a stochastic error model.

    \item  We provide extensive simulation results to validate the proposed framework. It is shown that RS-SCMA consistently outperforms conventional multicarrier RSMA and SCMA in terms of BER, block error rate (BLER), and sum-rate. The results further show that the splitting factor $\alpha$ serves as a control knob to strike a balance between the throughput and error-rate performance.
    \end{itemize}}
 \subsection{Notations}
In this paper, regular, bold lowercase, bold uppercase, and script fonts denote scalars, vectors, matrices, and sets, respectively. $\mathbb{C}^{J\times 1}$ represents a complex vector of dimension $J\times 1$. $\mathcal{CN}(\mu, \sigma^2)$ denotes a complex Gaussian distribution with mean $\mu$ and variance $\sigma^2$. The operators $(\cdot)^H$ and $(\cdot)^T$ represent the conjugate transpose and transpose operations, respectively. $\mathbf{I}_K$ denotes a $K \times K$ identity matrix. %The notation \( \| \mathbf{x} \|^2_{\mathbf{A}^{-1}} \) denotes the squared weighted norm of a vector $\mathbf{x}$ with respect to the matrix $\mathbf{A}$, defined as \( \mathbf{x}^H \mathbf{A}^{-1} \mathbf{x} \). 
The different notations and its meaning are given in TABLE~\ref{tab:notation_dims_rates}.
\begin{table}[hbpt]
\centering
\color{black}
\caption{\color{black} Symbols and notations.}
\label{tab:notation_dims_rates}
\setlength{\tabcolsep}{3pt}
\renewcommand{\arraystretch}{0.98}
\small
\begin{tabular}{p{0.20\columnwidth} p{0.76\columnwidth}}
\hline
\textbf{Symbol} & \textbf{Physical Meaning}\\
\hline
$J,K$ & Number of users and subcarriers. \\
$u,v$ & User indices. \\
$\mathbf{F}$ &Indicator/Factor-graph matrix $\mathbf{F}\in\{0,1\}^{K\times J}$\\ 
$d_f$ & Subcarrier node (SN) degree  $d_f=|\{j:F_{k,j}=1\}|$ (users per subcarrier) \\ 
$d_v$ & User node degree (UN) $d_v=|\{k:F_{k,j}=1\}|$ (subcarriers per user) \\ 
$N$ & Number of symbols per user. \\
$\alpha$ & RS factor; $l_c=\alpha N$, $l_p=(1-\alpha)N$.\\
$\lambda$ & Overloading factor.\\
$M_c$ & Common modulation order per subcarrier node.\\
$M_p$ & Codebook size, $|\mathcal{C}_v|=M_p$.\\
$\mathbf{H}_u,\hat{\mathbf{H}}_u$ & True/estimated channel ($K\times K$, usually diagonal).\\
$\mathbf{y}_j^{\text{diff}}$ & Post-SIC signal obtained at the receiver.\\
$\mathbf{y}_u,\mathbf{y}'_u$ & Pre-/post-SIC observations used in the sum-rate derivation ($K\times 1$).\\
$\mathbf{n}_u$ & Additive white Gaussian noise (AWGN), $\mathcal{CN}(\mathbf{0},N_0\mathbf{I}_K)$.\\
$\mathbf{s}_c,\mathcal{S}_c$ & Common vector and alphabet ($|\mathcal{S}_c|=M_c^K$ typically).\\
$\mathbf{c}_v,\mathcal{C}_v$ & Private SCMA codeword and codebook.\\
$p_c,p_p,p_{p,v}$ & Common/private powers; $\sum_v p_{p,v}=p_p$.\\
$\epsilon,\mathbf{s}_{\mathrm{res}}$ & SIC residue factor and residual common term.\\
$r_c,r_p$ & Low-density parity check (LDPC) code rates.\\
$\boldsymbol{\zeta}_{\mathrm{tot},u},\mathcal{Z}_{\mathrm{tot},u}$ & Total private interference set (common decoding).\\
$\boldsymbol{\zeta}_{u},\mathcal{Z}_{u}$ & Multi user interference (MUI) set (private decoding).\\
$P(\boldsymbol{\zeta})$ & Induced discrete probability mass function (PMF) over interference set.\\
$\Delta^{(c)}_{ab},\Delta^{(p)}_{ab}$ & Common/private distance metrics.\\
$R_c^{(\cdot)},R_{p,u}^{(\cdot)}$ & Common (min-user) and private rates.\\
%$\kappa$ & $\kappa=K\!\left(\frac{1}{\ln 2}-1\right)$.\\
\hline
\end{tabular}
%\vspace{-0.5em}
\end{table}

 \subsection{Organization}
The paper is organized as follows. Section~\ref{prel} introduces the basic concepts of RSMA and SCMA. Section~\ref{model} presents the proposed RS-SCMA system and is divided into six subsections. Section~\ref{Tx} gives details of the  RS-SCMA  transmitter, Section~\ref{Rx} gives details of the the RS-SCMA receiver (Rx-1). Section~\ref{ol_fact} discusses the calculation of the overloading factor for the proposed RS-SCMA systems and the spectral efficiency for different configurations of RS-SCMA architectures. Section~\ref{subsec:achievable_sum_rate} presents the mathematical derivation of RS-SCMA rate expression. Section~\ref{rx2} presents receiver (Rx-2) for coded RS-SCMA. Section~\ref{com_an} provides an insight into the complexity analysis of the RS-SCMA architecture. Section~\ref{result} discusses and presents the simulation results, Section~\ref{sum_rate} gives the sum rate plots for RS-SCMA compared to other systems, Section~\ref{ber} gives the BER plots for different configurations  and Section~\ref{conc} concludes the paper.

\section{Preliminaries}\label{prel}
In this section, the fundamentals of SCMA and RSMA systems are discussed as follows.
\subsection{A Brief Introduction of SCMA}
{\color{black}
SCMA is a code-domain NOMA scheme in which multiple users share certain number of orthogonal \emph{subcarriers} by transmitting multi-dimensional sparse codewords instead of scalar modulation symbols \cite{SCMA1}. Each user employs a predefined codebook that maps $\log_2M$ bits directly onto one of $M$ sparse $K$-dimensional codewords, where $K$ denotes the total number of subcarriers. We consider a system with $J$ users ($J>K$) multiplexed over $K$ subcarriers, thereby enabling overloaded transmission.

\subsubsection*{Indicator Matrix and Factor Graph}
The structure of an SCMA system is specified by a sparse indicator matrix $\mathbf{F} \in \{0,1\}^{K \times J}$. The $k$th row corresponds to subcarrier $k$, and the $j$th column corresponds to user $j$. ${\mathbf{F}}_{(k,j)}=1$ indicates that user $j$ occupies subcarrier $k$. The sparsity pattern governs both the multi-user interference and the receiver complexity.

Two key structural parameters are:
\begin{itemize}
    \item \textit{Subcarrier-node (SN) degree} $d_f$: the number of users sharing a subcarrier (non-zero entries per row).
    \item \textit{User-node (UN) degree} $d_v$: the number of subcarriers occupied by a user (non-zero entries per column).
\end{itemize}

The indicator matrix is typically designed to avoid short cycles (especially 4-cycles) to improve the convergence of message passing. An example for $J=6$, $K=4$, $d_v=2$, and $d_f=3$ is
\begin{equation}
\label{F_mat}
\mathbf{F}=
\begin{bmatrix}
1 & 0 & 1 & 0 & 1 & 0\\
0 & 1 & 1 & 0 & 0 & 1\\
1 & 0 & 0 & 1 & 0 & 1\\
0 & 1 & 0 & 1 & 1 & 0
\end{bmatrix}.
\end{equation}

\begin{figure}[!htb]
	\begin{center}
		\scalebox{0.9}{
			\begin{tikzpicture}[node distance=2.5cm,auto,>=latex']
			\draw [] (-1.2,0) node (xaxis) [] {User nodes};
			\draw [] (0,0) node (xaxis) [] { \large{$\bigcirc$}};
			\draw [] (0,0) node (xaxis) [] {$1$};	
			\draw [] (1,0) node (xaxis) [] {\large{$\bigcirc$}};\
			\draw [] (1,0) node (xaxis) [] {$2$};
			\draw [] (2,0) node (xaxis) [] {\large{$\bigcirc$}};
			\draw [] (2,0) node (xaxis) [] {$3$};	
			\draw [] (3,0) node (xaxis) [] {\large{$\bigcirc$}};
			\draw [] (3,0) node (xaxis) [] {$4$};
			\draw [] (4,0) node (xaxis) [] {\large{$\bigcirc$}};
			\draw [] (4,0) node (xaxis) [] {$5$};	
			\draw [] (5,0) node (xaxis) [] {\large{$\bigcirc$}};
			\draw [] (5,0) node (xaxis) [] {$6$};
			%%%%%
			\draw [] (-.9,-1.3) node (xaxis) [] {Subcarrier nodes};
			\draw  [] (0.7,-1.3) node (xaxis) [] {\Large{$\square$}};
			\draw  [] (0.7,-1.3) node (xaxis) [] {$1$};		
			\draw [] (1.9,-1.3) node (xaxis) [] {\Large{$\square$}};
			\draw [] (1.9,-1.3) node (xaxis) [] {$2$};
			\draw [] (3.1,-1.3) node (xaxis) [] {\Large{$\square$}};
			\draw [] (3.1,-1.3) node (xaxis) [] {$3$};
			\draw [] (4.4,-1.3) node (xaxis) [] {\Large{$\square$}};
			\draw [] (4.4,-1.3) node (xaxis) [] {$4$};

			\draw[-] (0,-0.17) -- (0.7,-1.13);
			\draw[-] (1,-0.17) -- (1.9,-1.13);
			\draw[-] (2,-0.17) -- (0.7,-1.13);

			\draw[-] (0,-0.17) -- (3.1,-1.13);
			\draw[-] (3,-0.17) -- (4.4,-1.13);
			\draw[-] (4,-0.17) -- (0.7,-1.13);

			\draw[-] (1,-0.17) -- (4.4,-1.13);
			\draw[-] (3,-0.17) -- (3.1,-1.13);
			\draw[-] (5,-0.17) -- (3.1,-1.13);

			\draw[-] (2,-0.17) -- (1.9,-1.13);
			\draw[-] (4,-0.17) -- (4.4,-1.13);
			\draw[-] (5,-0.17) -- (1.9,-1.13);

			\begin{pgfonlayer}{background}
			\path[fill=blue!20!white,rounded corners, draw=black!10] (-2,-1.8) -- (5.3,-1.8) -- (5.3,.4) -- (-2,.4) -- (-2,-1.8);
			\end{pgfonlayer}
			\end{tikzpicture}}
	\end{center}
	\caption{\footnotesize{Factor graph of $J=6$ users and $K=4$ subcarriers with $d_v=2$ and $d_f=3$.}}
	\label{factor_graph1}
\end{figure}

The corresponding bipartite graph/factor graph is shown in Fig.~\ref{factor_graph1}, where circles represent UNs and squares represent SNs. User~$j$ connects to the $d_v$ subcarriers indicated by the non-zero entries of column $j$ in $\mathbf{F}$.

\subsubsection*{Codebooks and Encoding}
Each user $j$ is assigned a codebook $\mathcal{C}_j \in \mathbb{C}^{K \times M}$,
\[
\mathcal{C}_j = \left[\mathbf{x}_{j1},\, \mathbf{x}_{j2},\, \ldots,\, \mathbf{x}_{jM}\right],
\]
where each codeword $\mathbf{x}_{jm}\in\mathbb{C}^{K}$ contains exactly $d_v$ non-zero entries at the subcarrier positions indicated by the $1$'s in column $j$ of $\mathbf{F}$. An example of a $4\times 6$ SCMA codebook set is shown in Fig.~\ref{scma_basic}.

\begin{figure}[hpbt]
\centering
\includegraphics[width=0.9\linewidth]{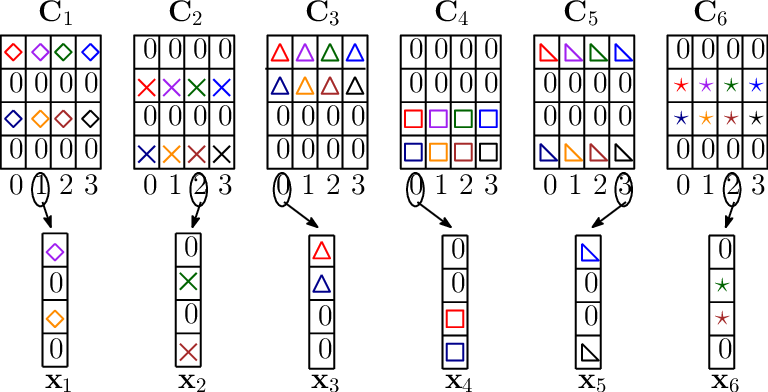}
\caption{\footnotesize Structure of an $6\times 4$ SCMA codebook set.}
\label{scma_basic}
\end{figure}

The overloading factor is defined as
\[
\lambda = \frac{J}{K}.
\]
For example in Fig.~\ref{factor_graph1}, $\lambda=1.5$ (i.e., $150\%$ overloading). 

To transmit information, user $j$ maps $\log_2(M)$ bits to an index $m \in \{1,\dots,M\}$ and transmits the corresponding sparse codeword $\mathbf{x}_{jm}$ taken from ${\cal{C}}_j$.

\subsubsection*{Received Signal Model}
In the downlink, the BS superimposes the SCMA codewords of all $J$ users over the $K$ subcarriers. The received signal at user $j$ can be expressed as
\begin{equation}
\label{eq:scma_received_signal}
\mathbf{y}_j = \mathbf{H}_j \left( \sum_{i=1}^{J} \mathbf{x}_i \right) + \mathbf{w}_j,
\end{equation}
where $\mathbf{H}_j=\mathrm{diag}(\mathbf{h}_j)\in\mathbb{C}^{K\times K}$ is the diagonal channel matrix with $\mathbf{h}_j=[h_{1,j},\ldots,h_{K,j}]^T$, and $\mathbf{w}_j \sim \mathcal{CN}(\mathbf{0},\sigma^2\mathbf{I}_K)$ denotes additive white Gaussian noise (AWGN).

\subsubsection*{SCMA Detection}
The sparsity of $\mathbf{F}$ yields a sparse factor graph that enables low-complexity multi-user detection using the MPA. UNs and SNs iteratively exchange likelihood messages, achieving near maximum \textit{a posteriori} (MAP) performance with substantially reduced complexity.
Due to space constraints, we omit the standard MPA update equations. Readers may refer to \cite{SCMA_det_low_complx,SCMA_det_low_complx1,Sharma2021} for detector design, and to \cite{scma_cb,Li2022,capacity_based_SCMA,Deka_DE} for multidimensional codebook constructions based on distance metrics, constellation shaping, capacity-based optimization, and differential evolution algorithms, respectively.
}

\begin{figure}[!hpbt]
		\centering
		\includegraphics[width=1\linewidth]{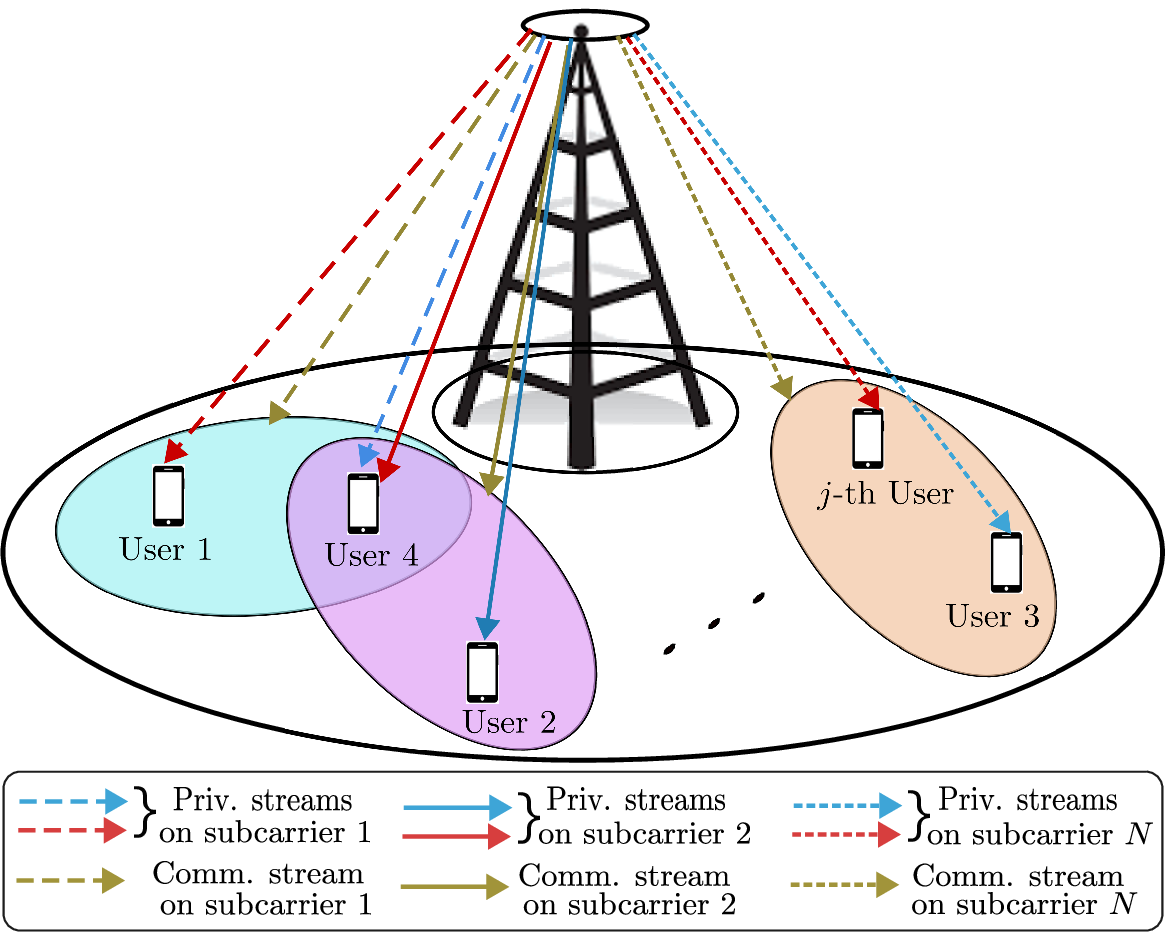}
		\caption{\footnotesize{Downlink MC-RSMA system.}}
		\label{rsma_basic}
	\end{figure}
{\color{black}\subsection{SISO Multicarrier RSMA (MC-RSMA)}
\label{subsec:mcrsma}

Consider a downlink SISO MC-RSMA system where a base station (BS) equipped with a single antenna serves a set of users $\mathcal{J}=\{1,\dots,J\}$ over $K$ orthogonal subcarriers $\mathcal{K}=\{1,\dots,K\}$. MC-RSMA \cite{mc_rsma,Pereira2025,Chen2021,Chen2020_mc_rsma} enables the BS to multiplex multiple users on the same frequency resource by superposing a common stream and multiple private streams in the power domain. As illustrated in Fig.~\ref{rsma_basic}, the BS applies this superposition on each subcarrier, and each receiver employs SIC to recover its intended message.

\begin{figure}[htpb!]
    \centering
    \includegraphics[width=1\linewidth]{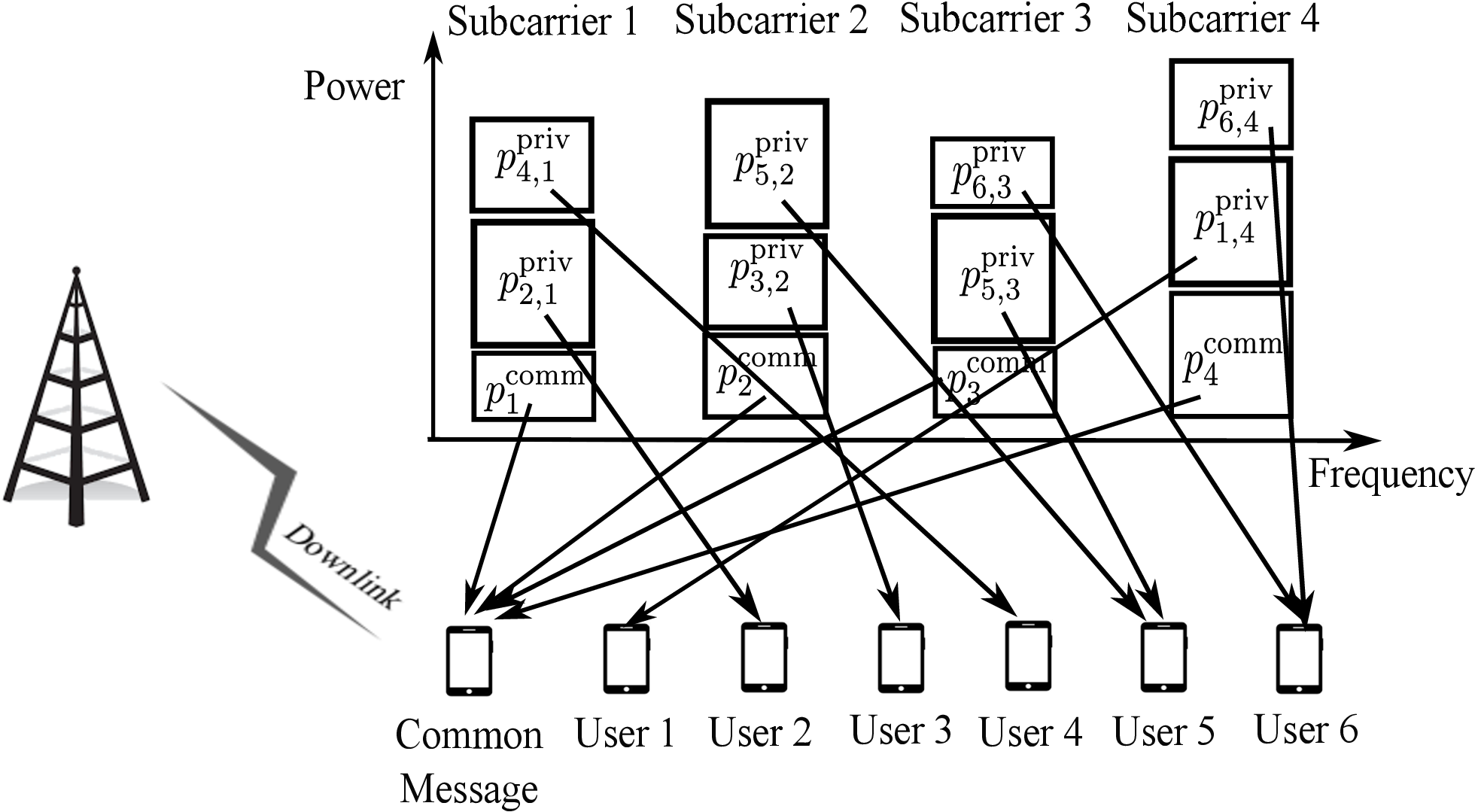}    \caption{\textcolor{black}{A typical user-subcarrier matching for a MC-RSMA downlink system.}}    \label{fig:mcrsma_ed}
\end{figure}
Let $\mathcal{J}_k \subseteq \mathcal{J}$ denote the subset of users co-scheduled on subcarrier $k\in\mathcal{K}$. The complex channel coefficient from the BS to user $j$ on subcarrier $k$ is denoted by $h_{j,k}\in\mathbb{C}$. User $j$ observes additive noise $z_{j,k}\sim\mathcal{CN}(0,\sigma^2)$.

% (Keep your system figure if needed; it is RS-SCMA-specific but fine to retain.)
\begin{figure*}[!htbp]
\centering
\includegraphics[width=0.95\textwidth]{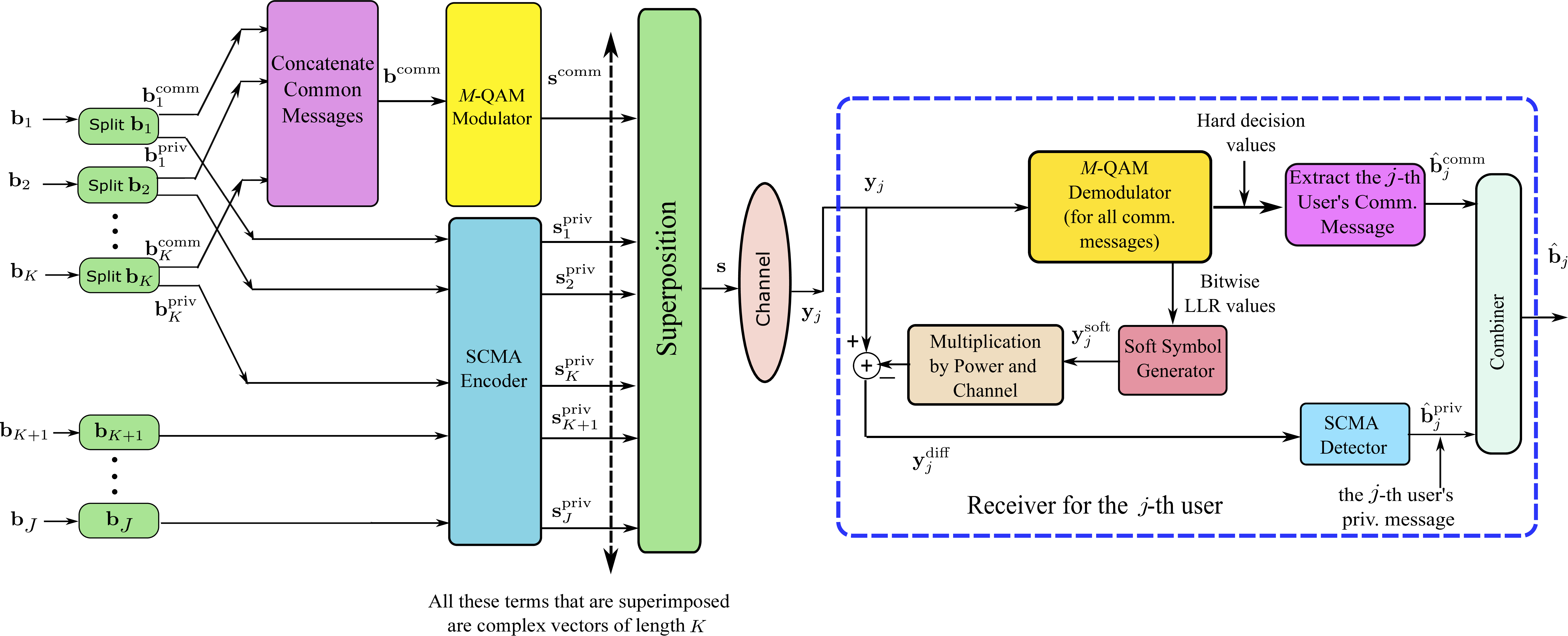}
\caption{\footnotesize{Proposed downlink RS-SCMA system architecture for $J$ users over $K$ subcarriers. The common messages are transmitted using $M$-QAM modulation, while private messages are SCMA-encoded. The receiver shown corresponds to Rx-1, which performs soft SIC using LLRs obtained directly from the QAM demodulator.}} 
\label{RSMA2_system1}
\end{figure*}

On each subcarrier $k\in\mathcal{K}$, the BS splits the messages of the users in $\mathcal{J}_k$ into common and private parts. The BS aggregates the common parts and encodes them into a single common stream $s_{c,k}$, while it encodes the private part of each user $j\in\mathcal{J}_k$ into the private stream $s_{j,k}$. Assuming normalized symbol energies $\mathbb{E}[|s_{c,k}|^2]=\mathbb{E}[|s_{j,k}|^2]=1$, the  signal transmitted by the BS  on subcarrier $k$ is given by
% \begin{equation}
% x_k = \sqrt{p_{c,k}}\, s_{c,k} + \sum_{j \in \mathcal{J}_k} \sqrt{p_{j,k}}\, s_{j,k},
% \label{eq:mcrsma_tx}
% \end{equation}
% --- Transmit signal on subcarrier k ---
\begin{equation}
x_k = \sqrt{p^{\mathrm{comm}}_{k}}\, s_{c,k}
      + \sum_{j \in \mathcal{J}_k} \sqrt{p^{\mathrm{priv}}_{j,k}}\, s_{j,k},
\label{eq:mcrsma_tx}
\end{equation}
where $p^{\mathrm{comm}}_{k}\ge 0$ and $p^{\mathrm{priv}}_{j,k}\ge 0$ denote the powers allocated to the common stream and user-$j$'s private stream on subcarrier $k$, respectively. Fig.~\ref{fig:mcrsma_ed} shows a generalized power distribution of an SISO MC-RSMA system with $J=6$ users and $K=4$ subcarriers.

MC-RSMA performs joint power allocation across subcarriers and streams. The power consumed on subcarrier $k$ is
$p_k \triangleq p^{\mathrm{comm}}_{k} + \sum_{j \in \mathcal{J}_k} p^{\mathrm{priv}}_{j,k},$ where the total transmit power is constrained by
% \begin{equation}
% \sum_{k=1}^{K} p_k \le P_{\mathrm{tot}}.
% \label{eq:total_power}
% \end{equation}
% --- Power per subcarrier and total power constraint ---
\begin{equation}
\sum_{k=1}^{K} p_k \le P_{\mathrm{tot}}.
\label{eq:total_power}
\end{equation}
%The split between the common and private layers controls the fraction of multi-user interference that users decode and cancel via SIC versus the residual interference treated as noise.

The received signal at user $j\in\mathcal{J}_k$ on subcarrier $k$ is
% \begin{equation}
% y_{j,k} = h_{j,k}\left(\sqrt{p_{c,k}}\, s_{c,k} + \sum_{\ell \in \mathcal{J}_k} \sqrt{p_{\ell,k}}\, s_{\ell,k}\right) + z_{j,k}.
% \label{eq:mcrsma_rx}
% \end{equation}
% --- Received signal at user j on subcarrier k ---
\begin{equation}
y_{j,k} = h_{j,k}\!\left(\sqrt{p^{\mathrm{comm}}_{k}}\, s_{c,k}
+ \sum_{\ell \in \mathcal{J}_k} \sqrt{p^{\mathrm{priv}}_{\ell,k}}\, s_{\ell,k}\right) + z_{j,k}.
\label{eq:mcrsma_rx}
\end{equation}
Following the one-layer RSMA protocol, user $j$ first decodes the common stream $s_{c,k}$ by treating all private streams as interference. The corresponding signal-to-interference-and-noise (SINR) for decoding the common stream at user $j$ is
% \begin{equation}
% \gamma^{(c)}_{j,k}
% = \frac{p_{c,k}|h_{j,k}|^2}{\sum_{\ell\in\mathcal{J}_k} p_{\ell,k}|h_{j,k}|^2 + \sigma^2}.
% \label{eq:sinr_common}
% \end{equation}
% --- SINR for decoding the common stream ---
\begin{equation}
\gamma^{(c)}_{j,k}
= \frac{p^{\mathrm{comm}}_{k}|h_{j,k}|^2}
{\sum_{\ell\in\mathcal{J}_k} p^{\mathrm{priv}}_{\ell,k}|h_{j,k}|^2 + \sigma^2}.
\label{eq:sinr_common}
\end{equation}

After decoding $s_{c,k}$, user $j$ reconstructs and subtracts the common component $h_{j,k}\sqrt{p_{c,k}}\, s_{c,k}$ via perfect SIC. User $j$ then decodes its private stream $s_{j,k}$ while treating the remaining private streams as noise. The resulting SINR for decoding the private stream is
% \begin{equation}
% \gamma^{(p)}_{j,k}
% = \frac{p_{j,k}|h_{j,k}|^2}{\sum_{\ell\in\mathcal{J}_k\setminus\{j\}} p_{\ell,k}|h_{j,k}|^2 + \sigma^2}.
% \label{eq:sinr_private}
% \end{equation}
% --- SINR for decoding user-j private stream after SIC ---
\begin{equation}
\gamma^{(p)}_{j,k}
= \frac{p^{\mathrm{priv}}_{j,k}|h_{j,k}|^2}
{\sum_{\ell\in\mathcal{J}_k\setminus\{j\}} p^{\mathrm{priv}}_{\ell,k}|h_{j,k}|^2 + \sigma^2}.
\label{eq:sinr_private}
\end{equation}
}

\section{Proposed RS-SCMA System Model}\label{model}
The proposed RS-SCMA system model combines rate-splitting and SCMA techniques to enhance system performance in terms of BER, sum-rate, overloading factor and spectral efficiency. The proposed system is considered for the downlink communication system.  In this work, bits refer to the raw binary data, whereas symbols denote the grouped bits mapped to modulation levels (e.g., QPSK symbol from 2 bits), which are then used for SCMA encoding or QAM modulation.

\subsection{Transmitter for RS-SCMA}\label{Tx} 

The proposed RS-SCMA transmitter architecture for downlink communication over $K$ shared subcarrier nodes serving $J$ users is illustrated in Fig.~\ref{RSMA2_system1}. In this framework, each user's message is split into a common and a private component. Here, symbols represent groups of bits (e.g., \{00, 01, 10, 11\}) that are indexed as \{1, 2, 3, 4\} depending on the modulation scheme. The common message is modulated using $M$-QAM (e.g., $M=4$ for QPSK), while the private message is encoded using SCMA.  The resulting signals are superimposed and transmitted over $K$ subcarrier nodes. The operation of the transmitter is further elaborated as follows:
\begin{comment}
The common message stream comprises parts of the messages intended to be decoded by all users, and is constructed by collecting selected symbols from the $K$ users. In contrast, private message components are intended for individual decoding by each respective user.

The RS-SCMA architecture integrates SCMA encoding and detection within the RSMA framework to improve spectral efficiency and enable user overloading ($J > K$). Message splitting proceeds as follows:
\begin{itemize}
  \item The common message symbols are modulated using $M$-QAM.
  \item The private message symbols are SCMA-encoded using user-specific codebooks (with $M=4$).
\end{itemize}
\end{comment}

Each user $j$ is assigned a symbol sequence $\mathbf{b}_j \in \{1, 2, \ldots, M\}^N$, consisting of $N$  sequence of discrete symbols. This sequence is partitioned into:
\begin{align}
  \mathbf{b}_j^{\text{comm}} &\in \{1, 2, \ldots, M\}^{l_c}, \\
  \mathbf{b}_j^{\text{priv}} &\in \{1, 2, \ldots, M\}^{l_p},
\end{align}
where the lengths $l_c$ and $l_p$ are determined by a splitting factor $\alpha \in [0, 1]$ as:
\begin{align}
  l_c = \alpha N, \quad l_p = (1 - \alpha) N,
  \label{lclp}
\end{align}
with $N$ and $\alpha$ chosen such that both $l_c$ and $l_p$ are integers.

For each channel use $n = 1, \ldots, l_c$, the BS forms the column vector:
\[
\mathbf{b}_n^{\text{comm}} =
\begin{bmatrix}
  b_{1,n}^{\text{comm}} \\
  b_{2,n}^{\text{comm}} \\
  \vdots \\
  b_{K,n}^{\text{comm}}
\end{bmatrix}
\in \{1,2,\ldots,M\}^{K \times 1},
\]
where $b_{j,n}^{\text{comm}}$ represents the $n$th symbol of the common message sequence from user $j$. This vector is modulated jointly using an $M$-QAM constellation to generate the common symbol $\mathbf{s}^{\text{comm}} \in \mathbb{C}^{K \times 1}$ for that channel use. For notational simplicity, the time index is omitted hereafter.

Simultaneously, the private parts from all $J$ users are processed independently for SCMA encoding. For users $1$ to $K$, private symbols are taken from $\mathbf{b}_j^{\text{priv}}$, while for users $K+1$ to $J$, the full message sequence $\mathbf{b}_j$ is treated as private. Each private stream is encoded using a user-specific SCMA codebook to generate a sparse codeword $\mathbf{s}_j^{\text{priv}} \in \mathbb{C}^{K \times 1}$.

The BS transmits the superimposed signal:
\begin{equation}
  \mathbf{s} = \sqrt{p_c} \mathbf{s}^{\text{comm}} + \sqrt{p_p} \sum_{j=1}^{J} \mathbf{s}_j^{\text{priv}},
  \label{trans}
\end{equation}
where $p_c$ and $p_p$ denote the power allocated to the common and private parts, respectively.

\subsubsection*{Special Cases}
%\vspace{-0.05in}
\begin{itemize}
    \item \textbf{Equal Splitting:} For \(\alpha = 0.5\), the symbols are equally split:
    \[
    l_c = l_p = 0.5N.
    \]
    
    \item \textbf{Unequal Splitting:} For unequal split, \(\alpha\) determines the relative proportions:
    \begin{itemize}
    \item If \(\alpha < 0.5\), then \(l_p > l_c\), indicating that more symbols are allocated to the private stream.
    \item If \(\alpha > 0.5\), then \(l_c > l_p\), indicating that more symbols are allocated to the common stream.
\end{itemize}
\end{itemize}

%This formulation offers flexibility in message allocation, allowing the system to adapt to varying requirements. The split parameter $\alpha$ thus acts as a tunable design control.
This formulation offers flexibility in symbol-level message allocation, allowing the system to adapt to varying communication requirements. It accommodates scenarios where private streams dominate (\(l_p > l_c\)) or where common streams are emphasized (\(l_c > l_p\)). The splitting parameter \( \alpha \) thus serves as a tunable design variable, enabling a balance between common and private message portions to optimize performance according to system objectives such as throughput, fairness, or decoding complexity.

\subsection{Receiver for RS-SCMA ($\text{Rx-1}$)}\label{Rx} 
The received signal for user $j$ is given by
\begin{equation}	
\mathbf{y}_j = \mathbf{H}_j \mathbf{s} + \mathbf{w}_j,
\label{rx_rsscma}
\end{equation}
where $\mathbf{y}_j\in \mathbb{C}^{K\times 1}$, $\mathbf{s}$ is the $K \times 1$ superimposed transmitted signal vector, and ${\mathbf{w}}_j\sim {\mathcal{CN}} \left( 0,\sigma^2\right)$ represents the AWGN for the $j$th user. Each diagonal element of $\mathbf{H}_j$ captures the flat-fading gain on the corresponding subcarrier for user $j$.

Prior works such as \cite{rsma_receiever} explore various SIC and non-SIC based receiver architectures for RSMA, while \cite{soft_SIC} discusses soft bit calculation using log-likelihod ratio (LLRs). However, these studies do not incorporate MPA-based detection within RSMA receivers. The proposed RS-SCMA receiver is the first to implement soft SIC with MPA for enhanced robustness.\\

Note: {\textit{Throughout this work, `soft bits' refer to the LLR of the bits obtained from the demodulator or decoder, while `soft symbols' refer to the complex-valued expectations of modulation symbols computed from these LLRs.}}\\

Fig.~\ref{RSMA2_system1} illustrates the architecture of the proposed soft SIC based receiver, Rx-1. The processing for the $j$-th user begins with the received $K \times 1$ signal vector $\mathbf{y}_j$. First, zero-forcing (ZF) equalization is applied to mitigate channel effects, yielding the equalized signal $\tilde{\mathbf{y}}_j$. The subsequent steps focus on demodulating the common stream to enable its cancellation.

%For each element $\tilde{y}_k$ of $\tilde{\mathbf{y}}_j$, the squared Euclidean distance to each symbol $s_i$ in the $M$-ary QAM constellation alphabet $\mathcal{S}$ is calculated as
For each element, processing begins by computing the squared Euclidean distance between each component of the equalized vector $\tilde{\mathbf{y}}_j$. Every symbol in the $M$-ary QAM constellation alphabet is denoted by $\mathcal{S}$. Let $\tilde y_{k_j}$ represent the $k$-th component of $\tilde{\mathbf{y}}_j$ for notational simplicity, the user index $j$ is hereafter dropped as the context is clear. The distance is calculated according to:
\begin{equation}
d_k(i) = |\tilde{y}_k - s_i|^2, \quad \forall s_i \in \mathcal{S}.
\label{eq:euclidean_distance_revised}
\end{equation}
These distances are used to compute the bit-wise LLRs with $m=\{0,\cdots,M-1\}$ as
\begin{equation}
\text{LLR}_{k,m} = \log \left( 
\frac{\sum\limits_{s_i \in \mathcal{S} \,:\, b_m(s_i)=0} \exp\left(-\frac{d_k(i)}{2\sigma^2}\right)}
{\sum\limits_{s_i \in \mathcal{S} \,:\, b_m(s_i)=1} \exp\left(-\frac{d_k(i)}{2\sigma^2}\right)} 
\right),\qquad 
\label{eq:llr_mqam_revised}
\end{equation}
where $b_m(s_i)$ denotes the $m$-th bit in the binary representation of $s_i$, and $\sigma^2$ is the post-equalization noise variance per dimension.  
A hard decision $\hat{\mathbf{b}}_j^{\text{comm}}$ for the common message can be obtained by selecting $\arg\min_i d_k(i)$ for each $k$. To enable SIC, the LLRs are converted into posterior bit probabilities as
\begin{equation}
    P(b_m=0) = \frac{1}{1 + e^{-\text{LLR}_{k,m}}}, \quad P(b_m=1) = 1 - P(b_m=0).
    \label{eq:bit_prob}
\end{equation}
The likelihood of each constellation symbol is then computed as
\begin{equation}
    P_k(s_i) = \prod_{m=1}^{\log_2 M} P(b_m = b_m(s_i)),
    \label{eq:sym_prob}
\end{equation}
and the corresponding soft estimate of the $k$-th common symbol is given by
\begin{equation}
\hat{s}_k = \sum_{s_i \in \mathcal{S}} P_k(s_i) \cdot s_i.
\label{soft_est}
\end{equation}
Collecting these soft estimates yields the vector $\mathbf{y}_j^{\text{soft}} = [\hat{s}_1, \ldots, \hat{s}_K]^T$.
The reconstructed common signal is passed through the user’s channel $\mathbf{H}_j$, scaled by $\sqrt{p_c}$, and subtracted from the received vector to perform soft SIC:
\begin{equation}
\mathbf{y}_j^{\text{diff}} = \mathbf{y}_j - \sqrt{p_c} \, \mathbf{H}_j \, \mathbf{y}_j^{\text{soft}}.
\label{eq:sic}
\end{equation}
The  signal $\mathbf{y}_j^{\text{diff}}$ is then processed by an MPA-based SCMA detector to estimate the private message $\hat{\mathbf{b}}_j^{\text{priv}}$. The complete estimated message is obtained by concatenating the common and private parts:
\begin{equation}
    \hat{\mathbf{b}}_j = \left[ \hat{\mathbf{b}}_j^{\text{comm}}, \ \hat{\mathbf{b}}_j^{\text{priv}} \right].
\end{equation}
\textbf{Algorithm~\ref{algo:rx1}} outlines the steps of the proposed receiver Rx-1.

\setcounter{equation}{22}
\begin{figure*}[b]
\hrule
\begin{align}
\lambda_{\textnormal{RS-SCMA}} &= \frac{\min\{l_c, l_p\}(K_c + K_p) \lambda_1 + (\max\{l_c, l_p\} - \min\{l_c, l_p\}) K_{\textnormal{dom}} \lambda_2}{\min\{l_c, l_p\}(K_c + K_p) + (\max\{l_c, l_p\} - \min\{l_c, l_p\}) K_{\textnormal{dom}}} \nonumber \\
&= \frac{\min\{\alpha, 1 - \alpha\}(K_c + K_p) \lambda_1 + |1 - 2\alpha| K_{\textnormal{dom}} \lambda_2}{\min\{\alpha, 1 - \alpha\} (K_c + K_p) + |1 - 2\alpha| K_{\textnormal{dom}}}.
\label{of}
\end{align}
\begin{center}
where \( K_{\text{dom}} = K_p \) if \( l_p > l_c \), and \( K_c \) if \( l_c > l_p \).
\end{center}
\end{figure*}

\setcounter{equation}{19}
\begin{algorithm}[!t]
\DontPrintSemicolon
\SetAlgoLined
\SetKwInput{KwInit}{Input}
\SetKwInput{KwOutput}{Output}
{
\KwInit{Received signal $\mathbf{y}_j$, channel matrix $\mathbf{H}_j$}
\KwOutput{Estimated message $\hat{\mathbf{b}}_j = [\hat{\mathbf{b}}_j^{\text{comm}}, \hat{\mathbf{b}}_j^{\text{priv}}]$}

\textbf{1. ZF Equalization:} Compute the equalized vector $\tilde{\mathbf{y}}_j = \text{ZF}(\mathbf{y}_j)$ \\

\textbf{2. Demodulation and Soft Estimation:} \\
\For{$k = 1$ \KwTo $K$}{
    Compute $d_k(i) = |\tilde{y}_k - s_i|^2$, $\forall s_i \in \mathcal{S}$ as in \eqref{eq:euclidean_distance_revised} \\
    Determine hard decision $\hat{b}_k^{\text{comm}} = \arg\min_i d_k(i)$ \\[2pt]
    \For{$m = 1$ to $\log_2 M$}{
        Compute $\text{LLR}_{k,m}$ using \eqref{eq:llr_mqam_revised} \\
        Obtain $P(b_m = 0)$ and $P(b_m = 1)$ from \eqref{eq:bit_prob}
    }
    Compute symbol probability $P_k(s_i)$ using \eqref{eq:sym_prob} \\
    Estimate soft symbol $\hat{s}_k$ using \eqref{soft_est}
}
Form vectors: $\hat{\mathbf{b}}_j^{\text{comm}} = [\hat{b}_1^{\text{comm}}, \ldots, \hat{b}_K^{\text{comm}}]$ and $\mathbf{y}_j^{\text{soft}} = [\hat{s}_1, \ldots, \hat{s}_K]^T$ \\

\textbf{3. Soft Interference Cancellation:} \\
Compute interference-suppressed signal: $\mathbf{y}_j^{\text{diff}} = \mathbf{y}_j - \sqrt{p_c} \, \mathbf{H}_j \, \hat{\mathbf{s}}$ using \eqref{eq:sic} \\

\textbf{4. SCMA Detection:} Apply MPA on $\mathbf{y}_j^{\text{diff}}$ to estimate $\hat{\mathbf{b}}_j^{\text{priv}}$ \\

\textbf{5. Message Reconstruction:} Combine the decoded parts \\
% \[
% \hat{\mathbf{b}}_j = [\hat{\mathbf{b}}_j^{\text{comm}}, \ \hat{\mathbf{b}}_j^{\text{priv}}]
% \]
\caption{Proposed Soft SIC Based Receiver (Rx-1)}
\label{algo:rx1}}
\end{algorithm}

\textbf{Remark 1:} We considered encoding both common and private messages using SCMA, but this increased system complexity. The superposition of codewords made it difficult to isolate the common message and perform effective SIC.

\textbf{Remark 2:} An alternative architecture using SCMA for the common message and QAM for the private messages was investigated but found to be ineffective. Effective SIC requires reliable decoding of the common message, but the MPA detector for the common SCMA stream is severely hampered by the dense, unstructured interference from the private QAM symbols. This leads to high initial error rates, causing higher error propagation during SIC that ultimately degrades private message detection. Our proposed architecture strategically avoids this by using a simpler modulation for the common part, ensuring its reliable cancellation and preserving a clean signal structure for the subsequent MPA detection stage.

{\color{black}\subsection{Overloading Factor and Spectral Efficiency Analysis}
\label{ol_fact}

The performance of RS-SCMA  is fundamentally characterized by its overloading and spectral efficiency, which jointly determine the system’s capacity and efficiency under subcarrier constraints.} Consider an RS-SCMA system with $J$ users and $K$ subcarriers. Let $K_p$ and $K_c$ denote the number of private and common symbols transmitted by all users per channel use. The overloading factor is tunable depending on the lengths \( l_c \) and \( l_p \). As the lengths of the common and private symbol streams vary, the system adjusts the overloading factor accordingly. If $l_c \neq l_p$, the system operates in two phases; otherwise, only joint transmission occurs. These two phases are explained below.

\textbf{Phase 1: Joint Transmission:}\
For \( \min(l_c, l_p) \) channel uses, both private and common symbols are transmitted simultaneously. The overloading factor in this phase is:
\[
\lambda_1 = \frac{K_c + K_p}{K}.
\]

\textbf{Phase 2: Dominant Stream Transmission:}\
For the remaining \( |l_c - l_p| \) channel uses, only the dominant stream is transmitted. The overloading factor in this phase is:
\[
\lambda_2 = 
\begin{cases}
\frac{K_p}{K}, & \text{if } l_p > l_c \\
\frac{K_c}{K}, & \text{if } l_c > l_p.
\end{cases}
\]
The effective overloading factor for RS-SCMA can be derived by combining contributions from the two phases:
\begin{equation}
\small
\begin{split}
\lambda_{\text{RS-SCMA}} = \frac{\lambda_1 \cdot \text{(Total number of symbols transmitted in Phase 1)}}
{\text{Total number of symbols transmitted across both phases}} \\
+ \frac{\lambda_2 \cdot \text{(Total number of symbols transmitted in Phase 2)}}
{\text{Total number of symbols transmitted across both phases}}.
\end{split}
\label{eq:overall_overloading}
\end{equation}

The effective overloading factor can be expressed using two cases:

\textbullet{} \textbf{\underline{Case-1:}} When $l_c > l_p$ (i.e., $\alpha > 0.5$)
\begin{align}
\lambda_{\text{RS-SCMA}} &= \frac{l_p (K_c + K_p) \lambda_1 + (l_c - l_p) K_c \lambda_2}{l_p (K_c + K_p) + (l_c - l_p) K_c}.\nonumber\\
\textnormal{In terms of }\alpha,\nonumber\\
\lambda_{\text{RS-SCMA}} &= \frac{(1 - \alpha)(K_c + K_p) \lambda_1 + (2\alpha - 1) K_c \lambda_2}{(1 - \alpha)(K_c + K_p) + (2\alpha - 1)K_c}.
\label{lc_more}
\end{align}

\textbullet{} \textbf{\underline{Case-2:}} When $l_p > l_c$ (i.e., $\alpha < 0.5$):
\begin{align}
\lambda_{\text{RS-SCMA}} &= \frac{l_c (K_c + K_p) \lambda_1 + (l_p - l_c) K_p \lambda_2}{l_c (K_c + K_p) + (l_p - l_c) K_p}.\nonumber\\
\textnormal{In terms of }\alpha,\nonumber\\
\lambda_{\text{RS-SCMA}} &= \frac{\alpha (K_c + K_p) \lambda_1 + (1 - 2\alpha)K_p \lambda_2}{\alpha (K_c + K_p) + (1 - 2\alpha)K_p}.
\label{lp_more}
\end{align}

Consequently, the effective overloading factor $\lambda_{\text{RS-SCMA}}$ can be expressed in a general form as given in \eqref{of}. When equal splitting occurs at $\alpha=0.5$, \eqref{of} reduces to $\lambda_{\text{RS-SCMA}}=\frac{K_c+K_p}{K}$.

\begin{example}{Overloading Factor Calculation:}
Consider an RS-SCMA system with $J = 6$ users and $K = 4$ subcarriers, where $l_p = 0.75N$ and $l_c = 0.25N$ (i.e., $\alpha = 0.25$). Among them, four users transmit both common and private symbols, and all six users transmit only private symbols. The number of transmitted symbols per channel use is $K_p = 6$ and $K_c = 4$.

To summarize, the transmission is divided into two phases:
\begin{itemize}
    \item \textbf{Phase 1}: For $\alpha N = 0.25 N$ channel uses, the system transmits $K_c = 4$ common symbols and $K_p = 6$ private symbols over $K = 4$ subcarriers. This results in an overloading factor of:
    \[ \lambda_1 = \frac{K_c + K_p}{K} = \frac{4 + 6}{4} = 250\%. \]
    \item \textbf{Phase 2}: For $(1 - 2\alpha)N = 0.5N$ channel uses, the system transmits $K_p = 6$ private symbols over $K = 4$ subcarriers, resulting in:
    \[ \lambda_2 = \frac{K_p}{K} = \frac{6}{4} = 150\%. \]
\end{itemize}

\textbf{Overall Overloading Factor:} Using \eqref{of}, we get:
\[ \lambda_{\textnormal{RS-SCMA}} = \frac{0.25 \cdot 10 \cdot 250\% + 0.5 \cdot 6 \cdot 150\%}{0.25 \cdot 10 + 0.5 \cdot 6} \approx 195.45\%. \]
\qedsymbol{}
\end{example}

{\color{black}

When the modulation orders for the common and private messages differ (i.e., $M_c \neq M_p$), the symbol-level overloading factor becomes an incomplete measure of data throughput. In such scenarios, spectral efficiency offers a more accurate metric. Spectral efficiency, denoted by $\eta$, quantifies the total number of information bits transmitted per subcarrier node in a single channel use. It is defined as:
\setcounter{equation}{23}
\begin{equation}
\eta = \frac{K_p \log_2 M_p + K_c \log_2 M_c}{K},
\end{equation}
where $K_p$ and $K_c$ denote the number of private and common symbols transmitted per channel use, respectively; $M_p$ and $M_c$ are their corresponding modulation orders and $K$ is the total number of subcarrier nodes.

As an illustrative example, consider the case of equal splitting with $\alpha = 0.5$ in our system setup, which results in $K_p = 6$ private symbols and $K_c = 4$ common symbols being multiplexed over $K=4$ subcarriers. We compare the spectral efficiency for two different modulation assignments:

\textbf{Case 1:} Private streams use SCMA codewords with \( M_p = 4 \), and common streams use QPSK (\( M_c = 4 \)):
\[
\eta_{\text{Case 1}} = \frac{6 \cdot 2 + 4 \cdot 2}{4} = \frac{20}{4} = 5 \text{ bits/subcarrier}.
\]

\textbf{Case 2:} Private streams use \( M_p = 4 \), and common streams use 8-QAM (\( M_c = 8 \)):
\[
\eta_{\text{Case 2}} = \frac{6 \cdot 2 + 4 \cdot 3}{4} = \frac{24}{4} = 6 \text{ bits/subcarrier}.
\]

In both cases, the overloading factor remains constant:
\[
\lambda_{\text{RS-SCMA}} = \frac{K_p + K_c}{K} = \frac{6 + 4}{4} = 2.5.
\]

This example demonstrates a key advantage of RS-SCMA: the system's spectral efficiency can be significantly enhanced by adapting the modulation order of one message layer (in this case, the common messages) without altering the symbol-level overloading. This highlights the inherent flexibility and scalability of the proposed framework.
}

\subsection{\color{black}Achievable Rate Analysis with Finite-Alphabet Constraints}
\label{subsec:achievable_sum_rate}
{\color{black}

%\subsubsection{Signal Model (Phase~1)}
Consider a downlink RS-SCMA system with $J$ users multiplexed over $K$ subcarriers. At user $u$, the received vector in the joint-transmission phase is 
\begin{equation}
\mathbf{y}_u
=
\mathbf{H}_u\!\left(
\sqrt{p_c}\,\mathbf{s}_c
+
\sum_{v=1}^{J}\sqrt{p_{p,v}}\,\mathbf{c}_v
\right)
+\mathbf{n}_u,
\label{eq:sys_rx_phase1}
\end{equation}
where $\mathbf{y}_u\in\mathbb{C}^{K\times 1}$, $\mathbf{H}_u\in\mathbb{C}^{K\times K}$, and $\mathbf{n}_u\in\mathbb{C}^{K\times 1}$.
The common-stream symbol vector is $\mathbf{s}_c\in\mathcal{S}_c\in\mathbb{C}^{K\times 1}$ with $|\mathcal{S}_c|=M_c^K$, and $\mathbf{c}_v\in\mathcal{C}_v\in\mathbb{C}^{K\times 1}$ is a sparse SCMA codeword of user $v$ with $|\mathcal{C}_v|=M_p$. 
The power coefficients satisfy $p_c\ge 0$ and $p_{p,v}\ge 0$, with $\mathbf{n}_u\sim\mathcal{CN}(\mathbf{0},N_0\mathbf{I}_K)$.%denotes the AWGN variance per complex dimension 

\subsubsection{Joint Transmission Rate (Phase 1) }
During this phase, which lasts for $\min\{l_c, l_p\}$ channel uses, both common and private streams are transmitted simultaneously.  The achievable rate for decoding in terms of the exact finite-alphabet and the approximated tractable lower bound are given as follows:

\subsubsection*{Exact Finite-Alphabet Rates}
Since the private-layer interference is discrete (finite-alphabet SCMA), the exact achievable rates follow the discrete-input continuous-output mutual information.
For user $u$, the aggregated private interference during common decoding is given  by
\begin{equation}
\boldsymbol{\zeta}_{\mathrm{tot},u}
\triangleq
\mathbf{H}_u\sum_{v=1}^{J}\sqrt{p_{p,v}}\,\mathbf{c}_v
\in\mathbb{C}^{K\times 1},
\label{eq:zeta_tot_def}
\end{equation}
where $\boldsymbol{\zeta}_{\mathrm{tot},u}$ can take values from the set of all possible aggregated interference realizations below
\begin{equation}
\mathcal{Z}_{\mathrm{tot},u}
\triangleq
\Big\{
\mathbf{H}_u\sum_{v=1}^{J}\sqrt{p_{p,v}}\,\mathbf{c}_v
~\big|~
\mathbf{c}_v\in\mathcal{C}_v,\ \forall v
\Big\}.
%\in\mathbb{C}^{K\times 1}.
\label{eq:Ztot_def_main}
\end{equation}
The detailed derivation  for individual users' common rate is given in the Appendix \ref{app:jensen_fa}.
The  exact effective common rate is limited by the worst user:
\begin{equation}
\begin{aligned}
R_c^{\mathrm{exact}} = \min_{u} \Bigg( & \log_2|\mathcal{S}_c| - \frac{1}{|\mathcal{S}_c|} \sum_{\mathbf{s}_a\in\mathcal{S}_c} \mathbb{E}_{\boldsymbol{\zeta}_{\mathrm{tot},u}} \bigg[ \\
& \mathbb{E}_{\mathbf{n}_u} \bigg\{ \log_2 \sum_{\mathbf{s}_b\in\mathcal{S}_c} \exp \left( -\frac{\Delta^{(c)}_{ab}}{N_0} \right) \bigg\} \bigg] \Bigg),
\end{aligned}
\label{eq:Rc_exact_main}
\end{equation}
where $\Delta^{(c)}_{ab}$ is the distance metric
\begin{equation}
\Delta^{(c)}_{ab}
=
\big\|
\sqrt{p_c}\mathbf{H}_u(\mathbf{s}_a-\mathbf{s}_b)
+\boldsymbol{\zeta}_{\mathrm{tot},u}
+\mathbf{n}_u
\big\|^2
-
\big\|
\boldsymbol{\zeta}_{\mathrm{tot},u}
+\mathbf{n}_u
\big\|^2 .
\label{eq:Delta_c_main}
\end{equation}

After decoding the common stream, user $u$ applies SIC. With SIC imperfection factor $\epsilon\in[0,1]$ is modeled as
\begin{equation}
\mathbf{y}'_u
=
\sqrt{p_{p,u}}\mathbf{H}_u\mathbf{c}_u
+\boldsymbol{\zeta}_{u}
+\sqrt{\epsilon p_c}\mathbf{H}_u\mathbf{s}_{c}
+\mathbf{n}_u,
\qquad
\mathbf{y}'_u\in\mathbb{C}^{K\times 1},
\label{eq:sys_rx_private_main}
\end{equation}
where $\boldsymbol{\zeta}_{u}\in\mathbb{C}^{K\times 1}$ is the private multi-user interference
\begin{equation}
\boldsymbol{\zeta}_{u}
\triangleq
\mathbf{H}_u\sum_{v\neq u}\sqrt{p_{p,v}}\,\mathbf{c}_v,
\label{eq:zeta_u_def}
\end{equation}
with the corresponding set of all possible realizations
\begin{equation}
\mathcal{Z}_{u}
\triangleq
\Big\{
\mathbf{H}_u\sum_{v\neq u}\sqrt{p_{p,v}}\,\mathbf{c}_v
~\big|~
\mathbf{c}_v\in\mathcal{C}_v,\ \forall v\neq u
\Big\}.
%\in\mathbb{C}^{K\times 1}
\label{eq:Zu_def_main}
\end{equation}
%and $\mathbf{s}_{\mathrm{res}}\in\mathcal{S}_c$ denotes the residual common symbol due to imperfect SIC. 
Assuming independent equiprobable codeword selection across users, the probability mass functions (PMFs) are uniform:
\begin{equation}
\begin{aligned}
P(\boldsymbol{\zeta}_{\mathrm{tot},u} = \mathbf{z}) &=
\begin{cases}
\dfrac{1}{|\mathcal{Z}_{\mathrm{tot},u}|}, & \mathbf{z} \in \mathcal{Z}_{\mathrm{tot},u}, \\[6pt]
0, & \text{otherwise},
\end{cases} \\[10pt]
P(\boldsymbol{\zeta}_{u} = \mathbf{z}) &=
\begin{cases}
\dfrac{1}{|\mathcal{Z}_{u}|}, & \mathbf{z} \in \mathcal{Z}_{u}, \\[6pt]
0, & \text{otherwise}.
\end{cases}
\end{aligned}
\label{eq:pmf_uniform_main}
\end{equation}

The exact private rate for user $u$ is
% \begin{equation}
% \begin{aligned}
% R_{p,u}^{\mathrm{exact}}
% =
% \log_2 M_p
% -\frac{1}{M_p}
% \sum_{\mathbf{c}_a\in\mathcal{C}_u}
% \mathbb{E}_{\boldsymbol{\zeta}_{u},\mathbf{s}_{\mathrm{res}}}
% \Big[
% \mathbb{E}_{\mathbf{n}_u}
% \Big\{
% \log_2
% \sum_{\mathbf{c}_b\in\mathcal{C}_u}
% \exp\!\Big(
% -\frac{\Delta^{(p)}_{ab}}{N_0}
% \Big)
% \Big\}
% \Big],
% \end{aligned}
% \label{eq:Rp_exact_main}
% \end{equation}
\begin{equation}
\begin{aligned}
R_{p,u}^{\mathrm{exact}} = & \log_2 M_p - \frac{1}{M_p} \sum_{\mathbf{c}_a\in\mathcal{C}_u} \mathbb{E}_{\boldsymbol{\zeta}_{u},\mathbf{s}_{\mathrm{c}}} \bigg[ \\
& \mathbb{E}_{\mathbf{n}_u} \Big\{ \log_2 \sum_{\mathbf{c}_b\in\mathcal{C}_u} \exp \Big( -\frac{\Delta^{(p)}_{ab}}{N_0} \Big) \Big\} \bigg],
\end{aligned}
\label{eq:Rp_exact_main}
\end{equation}

with $\Delta^{(p)}_{ab}$ given by
\begin{align}
\Delta^{(p)}_{ab}
=
&\big\|
\sqrt{p_{p,u}}\mathbf{H}_u(\mathbf{c}_a-\mathbf{c}_b)
+\boldsymbol{\zeta}_{u}
+\sqrt{\epsilon p_c}\mathbf{H}_u\mathbf{s}_{c}
+\mathbf{n}_u
\big\|^2
-\nonumber\\
&\big\|
\boldsymbol{\zeta}_{u}
+\sqrt{\epsilon p_c}\mathbf{H}_u\mathbf{s}_{c}
+\mathbf{n}_u
\big\|^2 .
\label{eq:Delta_p_main}
\end{align}

\subsubsection*{Tractable  Lower Bound}
The exact expressions in \eqref{eq:Rc_exact_main} and \eqref{eq:Rp_exact_main} are complicated because of the expectation over noise.
To obtain a tractable lower bound expression, we follow the standard finite-alphabet analysis in \cite{zeng_low, Wu, rsma_receiever}.
It results in a constant gap of  $\kappa = K\left(\frac{1}{\ln 2}-1\right)$.
The derivation of the steps are  described in Appendix~\ref{app:jensen_fa}.
The resulting tractable lower bounds are given as follows:
\begin{equation}
\begin{aligned}
R_{c,u}^{\mathrm{LB}}
&\approx
\log_2|\mathcal{S}_c|
-\frac{1}{|\mathcal{S}_c|}
\sum_{\mathbf{s}_a\in\mathcal{S}_c}
\log_2\!\Bigg(
\sum_{\mathbf{s}_b\in\mathcal{S}_c}
\sum_{\boldsymbol{\zeta}_{\mathrm{tot},u}}
P(\boldsymbol{\zeta}_{\mathrm{tot},u})
\\[-0.25em]
&\hspace{4.2em}\times
\exp\!\Bigg(
-\frac{\big\|
\sqrt{p_c}\mathbf{H}_u(\mathbf{s}_a-\mathbf{s}_b)
+\boldsymbol{\zeta}_{\mathrm{tot},u}
\big\|^2}{2N_0}
\Bigg)
\Bigg)
-\kappa,
\end{aligned}
\label{eq:Rc_LB_main}
\end{equation}

\begin{equation}
\begin{aligned}
&R_{p,u}^{\mathrm{LB}}
\approx
\log_2 M_p
-\frac{1}{M_p}
\sum_{\mathbf{c}_a\in\mathcal{C}_u}
\log_2\!\Bigg(
\sum_{\mathbf{c}_b\in\mathcal{C}_u}
\sum_{\boldsymbol{\zeta}_{u}}
\sum_{\mathbf{s}_{c}\in\mathcal{S}_c}
P(\boldsymbol{\zeta}_u)P(\mathbf{s}_{c})
\\[-0.05em]
&\hspace{0.5em}\times
\exp\!\Bigg(
-\frac{\big\|
\sqrt{p_{p,u}}\mathbf{H}_u(\mathbf{c}_a-\mathbf{c}_b)
+\boldsymbol{\zeta}_{u}
+\sqrt{\epsilon p_c}\mathbf{H}_u\mathbf{s}_{c}
\big\|^2}{2N_0}
\Bigg)
\Bigg)
-\kappa.
\end{aligned}
\label{eq:Rp_LB_main}
\end{equation}
The lower bound on the common rate is $R_c^{\mathrm{LB}}=\min_u R_{c,u}^{\mathrm{LB}}$. The Phase~1 sum rate (exact or lower bound) is
\begin{equation}
R^{(P1)}_{\mathrm{sum}}
=
R_c^{(\cdot)}
+\sum_{u=1}^{J}R_{p,u}^{(\cdot)},
\label{eq:sumrate_p1_main}
\end{equation}
where $(\cdot)$ denotes either $\mathrm{exact}$ or lower bound.

%-----------------------------
% Phase 2 (kept mostly text, but column-safe equations)
%-----------------------------
\subsubsection{Dominant Stream Phase (Phase~2)}
During Phase~2 (lasting $|l_c-l_p|$ channel uses), only the dominant stream is transmitted. Phase~2 follows from the same finite-alphabet formulation by setting the absent-stream power to zero:

\paragraph{Case 1: $l_p > l_c$ (i.e., $\alpha < 0.5$) (private-only transmission)}
The Phase~2 sum-rate is
\begin{equation}
R^{(\mathrm{P2,priv})}=\sum_{u=1}^{J} R_{p,u}^{(P2)},
\end{equation}
obtained from \eqref{eq:Rp_exact_main} (or \eqref{eq:Rp_LB_main}) by setting $\epsilon=0$ and removing the common layer.
\paragraph{Case 2: $l_c > l_p$ (i.e., $\alpha > 0.5$) common-only transmission}
The Phase~2 sum-rate is
\begin{equation}
R^{(\mathrm{P2,comm})}=\min_{u} R_{c,u}^{(P2)},
\end{equation}
obtained from \eqref{eq:Rc_exact_main} (or \eqref{eq:Rc_LB_main}) by setting $p_{p,v}=0$ for all $v$.

Thus,
\begin{equation}
R^{(P2)}=
\begin{cases}
R^{(\mathrm{P2,priv})}, & \alpha<0.5,\\
0, & \alpha=0.5,\\
R^{(\mathrm{P2,comm})}, & \alpha>0.5.
\end{cases}
\label{eq:P2_piecewise}
\end{equation}
\subsubsection*{Overall Achievable Sum-Rate}
The overall achievable sum-rate depends on the splitting factor $\alpha$ through the durations of the two phases.
Phase~1 (joint transmission) occupies  $\min\{l_c, l_p\}$ channel uses and achieves $R^{(P1)}$, whereas Phase~2 (dominant-stream transmission) occupies $|l_c-l_p|$ channel uses and achieves $R^{(P2)}$.
Hence, the total number of bits delivered over one RS-SCMA block equals
$\min\{l_c,l_p\}\,R^{(P1)} + |l_c-l_p|\,R^{(P2)}$.
Normalizing by the total active transmission duration $\max\{l_c,l_p\}$ gives the average sum-rate per channel use:
\begin{equation}
\label{eq:unified_final_rate}
R_{\text{RS-SCMA}}(\alpha)
=
\frac{\min\{\alpha,1-\alpha\}\,R^{(P1)} + |1-2\alpha|\,R^{(P2)}}{\max\{\alpha,1-\alpha\}}.
\end{equation}

The expression in \eqref{eq:unified_final_rate} is therefore a weighted time-average of the phase-wise sum-rates.
It reduces to $R^{(P1)}$ at $\alpha=0.5$ (no Phase~2), and to the corresponding single-stream rate at the extremes:
$R^{(\mathrm{P2,priv})}$ for $\alpha=0$ and $R^{(\mathrm{P2,comm})}$ for $\alpha=1$.

% \subsubsection{Overall Achievable Sum-Rate}
% The time-averaged achievable sum-rate is
% \begin{equation}
% R_{\text{RS-SCMA}}(\alpha)
% =
% \frac{\min(\alpha,1-\alpha)\,R^{(P1)} + |1-2\alpha|\,R^{(P2)}}
% {\max(\alpha,1-\alpha)}.
% \label{eq:unified_final_rate}
% \end{equation}
% This expression reduces to $R^{(P1)}$ at $\alpha=0.5$, to pure-SCMA (private-only) at $\alpha=0$, and to common-only transmission at $\alpha=1$.

% \subsubsection{Overall Achievable Sum-Rate}
% The overall achievable sum-rate, $R_{\text{RS-SCMA}}(\alpha)$, is defined as the time-averaged throughput over the total active transmission duration, $T = \max(l_c, l_p)$. By aggregating the bits transmitted during both Phase 1 and Phase 2 and normalizing by $T$, we obtain the unified expression:
% \begin{equation}
% 	\label{eq:unified_final_rate}
% 	R_{\text{RS-SCMA}}(\alpha) = \frac{\min(\alpha, 1-\alpha) R^{(P1)} + |1 - 2\alpha| R^{(P2)}}{\max(\alpha, 1-\alpha)}.
% \end{equation}
% This expression characterizes the system's performance across the full range of the splitting factor $\alpha$. Specifically, at $\alpha=0.5$ (where $l_c = l_p$), the rate reduces to the joint transmission sum-rate $R^{(P1)}$. At the boundaries, the expression recovers the rates of the standalone configurations: pure SCMA transmission ($R^{(P2, \text{priv})}$) at $\alpha=0$, and common-only transmission ($R^{(P2, \text{comm})}$) at $\alpha=1$.
}

\subsection{Coded Receiver (Rx-2):}
\label{rx2}

\begin{figure}[!hpbt]
    \centering
    \includegraphics[width=\linewidth]{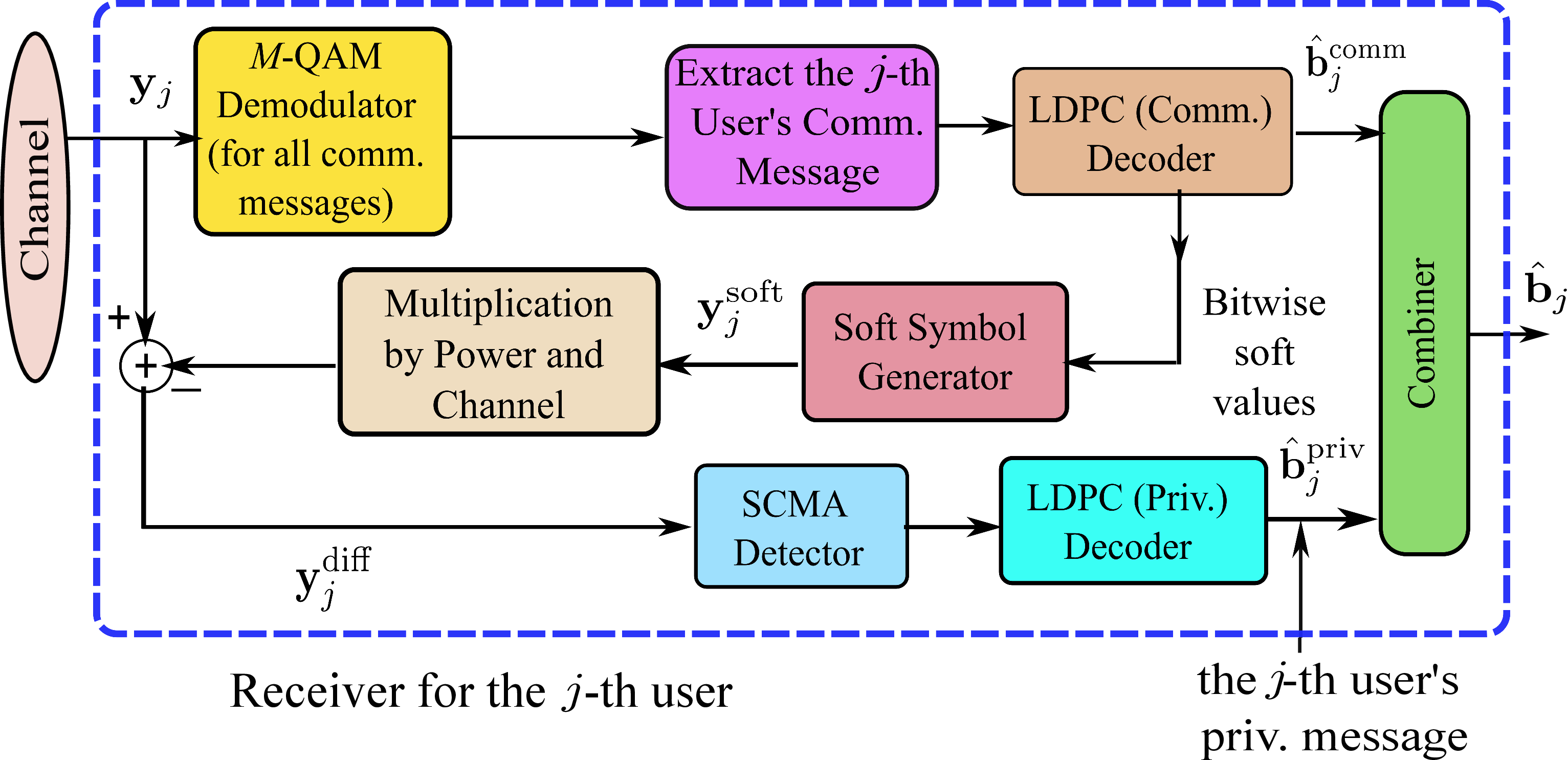}
    \caption{\footnotesize{Rx-2 architecture for LDPC-coded RS-SCMA, using LDPC decoder output for soft SIC.}}
    \label{RSMA2_rec}
\end{figure}

Rx-2 is a performance-enhanced receiver architecture designed for coded RS-SCMA systems, as shown in Fig.~\ref{RSMA2_rec}. Although Fig.~\ref{RSMA2_rec} illustrates LDPC-coded RS-SCMA, the framework supports any error-correcting code (e.g., LDPC, Turbo, Polar etc.). At the transmitter, each user’s message is split into common and private parts, encoded with code rates \( r_c \) and \( r_p \), respectively. Rx-1, shown in Fig.~\ref{RSMA2_system1}, was originally proposed for uncoded RS-SCMA. For coded scenarios, its structure can be extended by deriving soft bits for SIC directly from the QAM demodulator output and appending the channel decoder blocks at separate designated places. This coded Rx-1 setup relies on raw LLRs to generate soft symbols for common stream cancellation during SIC. However, due to decoding uncertainty, the interference reconstruction may be less accurate.
Rx-2 mitigates this by extracting soft bits from the output of the common stream’s channel decoder. These decoded probabilities yield more reliable soft symbols, enabling more accurate reconstruction and subtraction of the common signal via soft SIC. This reduces error propagation and enhances private stream detection. Specifically, the received signal \( \mathbf{y}_j \) undergoes ZF equalization and QAM demodulation to compute initial LLRs. These LLRs are passed to the channel decoder, which produces refined bitwise probabilities over the entire codeword. These are used to compute soft symbols as 
$\hat{s}_k = \sum_{i=1}^{M} P_k(s_i) \cdot s_i, \quad \text{for } k = 1,\ldots,K,$
where \( P_k(s_i) \) is the channel decoder-generated probability that the \( k \)-th symbol equals \( s_i \) computed as per \eqref{eq:sym_prob}. Observe that  Rx-2 get its soft symbols for soft SIC from the channel decoder, while coded Rx-1 obtains the same from the QAM demodulator. We form the vector $\mathbf{y}_j^{\text{soft}} = [\hat{s}_1, \ldots, \hat{s}_K]^T$. The reconstructed soft symbol vector \( \mathbf{y}_j^{\text{soft}} \) is then used for soft SIC:
\begin{equation}
\mathbf{y}_j^{\text{diff}} = \mathbf{y}_j - \sqrt{p_c} \, \mathbf{H}_j \mathbf{y}_j^{\text{soft}}.
\end{equation}

The resulting signal \( \mathbf{y}_j^{\text{diff}} \) exhibits reduced interference from the common stream and is fed to the SCMA based MPA detector, which generates soft LLRs of the private stream. These are finally decoded using the corresponding channel decoder. Compared to coded Rx-1, Rx-2 achieves lower BER due to cleaner interference cancellation and improved private stream detection.

\subsection{Complexity Analysis}
\label{com_an}
{
The detection process in the RS-SCMA receiver begins with the demodulation of the common message. For a $K$-dimensional received vector where each component is an $M_c$-ary QAM symbol, the demodulation complexity is linear with respect to both the number of subcarriers and the constellation size, given by $O(K M_c)$.

The core of a conventional SCMA receiver, and the second stage of our RS-SCMA receiver, is the MPA detector. The complexity of the MPA is dominated by the processing at the subcarrier nodes $d_f$. The complexity is $O(M_p^{d_f})$ operations per subcarrier node, where $M_p$ is the size of the SCMA codebook alphabet. For a system with $K$ subcarrier nodes, the total complexity of a standalone SCMA receiver is $\mathcal{X}_{\text{SCMA}} = O\left(K M_p^{d_f}\right)$.
% \[
% \mathcal{C}_{\text{SCMA}} = O\left(K M_p^{d_v}\right).
% \]

The proposed RS-SCMA receiver combines these operations with a low-complexity soft SIC step. The total complexity is the sum of the $M$-QAM demodulation ($O(K M_c)$), the soft symbol reconstruction and subtraction for SIC ($O(K)$), and the MPA-based SCMA detection ($O(K M_p^{d_f})$). Thus, the total complexity for the uncoded RS-SCMA case is $\mathcal{X}_{\text{RS-SCMA}} = O\left(K \left( M_c + 1 + M_p^{d_f} \right)\right)$.
% \[
% \mathcal{C}_{\text{RS-SCMA}} = O\left(K \left( M_c + 1 + M_p^{d_v} \right)\right).
% \]

Comparing the proposed scheme to the conventional SCMA benchmark, the complexity ratio is:
% \begin{equation}
% \label{eq:comp_ratio}
% \frac{\mathcal{C}_{\text{RS-SCMA}}}{\mathcal{C}_{\text{SCMA}}} \approx \frac{M_c + 1 + M_p^{d_v}}{M_p^{d_v}}.
% \end{equation}
\begin{equation}
\label{eq:comp_ratio}
\frac{\mathcal{X}_{\text{RS-SCMA}}}{\mathcal{X}_{\text{SCMA}}}
\approx
\frac{M_c + 1 + M_p^{d_f}}{M_p^{d_f}},
\end{equation}
where $\mathcal{X}_{(\cdot)}$ denotes the computational complexity (operation count).
This ratio highlights the trade-off between performance and complexity. For the system parameters  ($M_p = 4$, $d_f = 3$), using QPSK for the common message ($M_c=4$) results in a complexity ratio of $1.078$, a mere 7.8\% increase over conventional SCMA. If the common message modulation is upgraded to 8-QAM ($M_c=8$) to boost throughput, the ratio becomes $1.203$ (a 20.3\% increase).

This modest increase in complexity yields a significant gain in spectral efficiency. For the configurations above, the spectral efficiencies are $5$~bits/subcarrier for $(M_p, M_c) = (4, 4)$ and $6$~bits/subcarrier use for $(M_p, M_c) = (4, 8)$. This demonstrates that increasing $M_c$ from 4 to 8 improves spectral efficiency by $20\%$ while incurring only a marginal $\approx 11.6\%$ additional complexity relative to the $M_c = 4$ case.

For coded RS-SCMA systems, the overall complexity also includes the channel decoding stage. Both coded Rx-1 and Rx-2 exhibit the same fundamental computational complexity, with the primary distinction being the source of soft information for SIC. While coded Rx-1 uses soft bits directly from the QAM demodulator, Rx-2 enhances SIC accuracy and overall BER performance by using more reliable soft bits extracted from the output of the common message's channel decoder. The complexity of a channel decoder can be abstracted as $\mathcal{X}_{\text{dec}}(n, r, \mathcal{A})$, where $n$ is the block length, $r$ is the code rate, and $\mathcal{A}$ represents algorithm-specific parameters. For instance, in some of our simulations, we have used LDPC codes. LDPC  decoder uses the belief propagation (BP) algorithm \cite{BP}, which incurs complexity \( O(I \cdot n \cdot d) \), where \( I \) is the number of iterations, \( n \) is the codeword length, and \( d \) is the average number of non-zero entries per column in the parity-check matrix. Since RS-SCMA performs decoding separately for the common and private streams, the total complexity includes both decoders, TABLE~\ref{tab:complexity_comparison} summarizes the complexity analysis. }
\begin{table}[!htbp] {
\caption{Complexity analysis of coded RS-SCMA receivers.}
\label{tab:complexity_comparison}
\centering
\small % reduced font
\begin{tabular}{|p{4.8cm}|p{2.5cm}|} % reduce column widths a bit
\hline
\textbf{Component} & \textbf{Complexity (\( O \))} \\ \hline
\( M_c \)-QAM Demodulation (Common Stream) & \( K M_c \) \\ \hline
Soft Bits Computation & \( K \) \\ \hline
Soft Symbol Generation & \( K \) \\ \hline
Soft SIC & \( K \) \\ \hline
SCMA Detection (Private Stream) & \( K M_p^{d_f} \) \\ \hline
Decoder (Common Stream) & \( \mathcal{X}_{\text{dec}}(n_c, r_c, \mathcal{A}_c) \) \\ \hline
Decoder (Private Stream) & \( \mathcal{X}_{\text{dec}}(n_p, r_p, \mathcal{A}_p) \) \\ \hline
\end{tabular}}
\end{table}

\setcounter{equation}{21}
\vspace{-0.1in}
\section{Simulation Results}\label{result}
In this section, the sum-rate and error-rate performance of the proposed downlink SISO RS-SCMA system is evaluated and compared against conventional SCMA and SISO MC-RSMA systems. All simulations are carried out in MATLAB R2023b and later versions. Unless otherwise specified, the channel model is Rayleigh fading, and the splitting parameter is set to $\alpha=0.5$ (equal splitting). The RS-SCMA configuration considers $J=6$ users and $K=4$ subcarriers, resulting in a maximum overloading factor of $\lambda_{\text{RS-SCMA}}=250\%$ as per \eqref{of} . For the coded RS-SCMA system, 5G-NR-based LDPC codes are employed \cite{ldpc_5g}. The power allocation to the common and private messages follows the max-min fairness (MMF) scheme \cite{mmf_noma}. The corresponding common-stream power allocation factors are: $p_c=0.9098$ at 0~dB, $p_c=0.9114$ at 5~dB, $p_c=0.9156$ at 10~dB, $p_c=0.9252$ at 15~dB, $p_c=0.9409$ at 20~dB, $p_c=0.9585$ at 25~dB, and $p_c=0.9734$ at 30~dB. Although RS-SCMA can support any $M$-QAM modulation for common messages, QPSK is employed here for demonstration. For a fair comparison, a $250\%$ overloaded SCMA system with $J=15$ users and $K=6$ subcarriers is considered. The SCMA codebook is designed via differential evolution \cite{Deka_DE}, using the following indicator matrix:
    \begin{align*}
    {\bf{F}}=
    \scalebox{0.8}{$
    \begin{bmatrix}
      1 & 0 & 0 & 1 & 0 & 0 & 1 & 0 & 0 & 1 & 0 & 0 & 1 & 0 & 0\\
      1 & 0 & 0 & 0 & 1 & 0 & 0 & 1 & 0 & 0 & 1 & 0 & 1 & 0 & 0\\
      0 & 1 & 0 & 1 & 0 & 1 & 0 & 0 & 1 & 0 & 0 & 0 & 0 & 1 & 0\\
      0 & 1 & 1 & 0 & 0 & 0 & 0 & 0 & 0 & 1 & 0 & 1 & 0 & 1 & 0\\
      0 & 0 & 0 & 0 & 1 & 0 & 1 & 0 & 1 & 0 & 1 & 0 & 0 & 0 & 1\\
      0 & 0 & 1 & 0 & 0 & 1 & 0 & 1 & 0 & 0 & 0 & 1 & 0 & 0 & 1\\
    \end{bmatrix}$}.
    \end{align*}
{\color{black}For a fair comparison, identical to the proposed SISO RS-SCMA system, we consider a SISO MC-RSMA with identical bandwidth as $K=4$ orthogonal subcarriers, identical user load as $J=6$ users, identical effective overloading factor: $6$ private symbols and $4$ common symbols over $4$ subcarriers ($\lambda_\text{MC-RSMA}=250\%$ overloading factor). The user-to-subcarrier mapping follows the user-pairing method proposed in \cite{Pereira2025}. This replaces the fixed SCMA factor graph with dynamic frequency-domain pairing. The SISO MC-RSMA benchmark adopts a common transmission structure and MMF power allocation adopted in \cite{Chen2020_mc_rsma, Chen2021}.}

%The simulation results are presented in two parts: sum-rate analysis and error-rate analysis of RS-SCMA under various configurations, including coded transmission with LDPC. Unless otherwise specified, all simulations are conducted in a Rayleigh fading channel environment and equal splitting with  $\alpha=0.5$.

  %However, a fundamental architectural difference must be highlighted, which underscores a key practical advantage of our approach. %The RSMA benchmark, being a MISO system, uses channel state information at the transmitter (CSIT) to design its precoders. In contrast, our proposed RS-SCMA is modeled as a system where detection relies only on CSIR. This comparison therefore demonstrates that our RS-SCMA framework achieves highly competitive performance without the significant overhead and feedback requirements associated with CSIT.
\subsection{Sum-Rate Analysis}\label{sum_rate}
{\color{black}
We begin the simulation results by comparing the achievable sum-rate of the proposed RS-SCMA against the SISO-SCMA and SISO MC-RSMA benchmarks.
To compare the spectral efficiency, we employ a consistent modulation order of $M=4$ for all transmitted streams. For the SISO MC-RSMA benchmark, we evaluate the sum-rate using a Gaussian approximation, as the finite-constellation capacity for this specific architecture remains an open research topic. However, in case of SISO MC-RSMA, the finite alphabet will give a sum-rate lower than Gaussian. Conversely, the SCMA benchmark utilizes codebooks of order $M=4$ optimized in \cite{capacity_based_SCMA}. Fig.~\ref{fig:sum_rate_r1} presents the sum-rate performance of the proposed RS-SCMA framework (configured at $\alpha=0.5$) against these benchmarks. We plot the RS-SCMA rates using the exact expression.
The expectation is evaluated via Monte Carlo simulation with $10^5$
 samples.
We also plot the approximated tractable lower bound derived for the sum-rate expression in \eqref{eq:sumrate_p1_main}.

In the low-to-mid SNR regime, RS-SCMA exhibits robust performance that aligns closely with conventional SCMA. By exploiting the structured sparsity of the underlying SCMA layer, the hybrid scheme effectively mitigates interference, yielding a distinct advantage over MC-RSMA in noise-limited scenarios.
At high SNRs, the structural limitations of conventional SCMA become evident as the sum-rate saturates near 12 bits/channel-use. This hard capacity ceiling results from the fixed overloading factor and the finite modulation order. RS-SCMA effectively circumvents this saturation. By treating part of the interference as a decodable common message, the rate-splitting component utilizes the improving channel conditions to push the throughput beyond the SCMA ceiling. Consequently, RS-SCMA combines the overload gain of SCMA with the interference management of RSMA. Furthermore, the results validate the tightness of the proposed approximation; the approximated lower bound tracks the exact RS-SCMA rate closely across the entire SNR range and converges asymptotically, confirming its utility for tractable system optimization.}
 \begin{figure}[!htpb]
    \centering
    \begin{tikzpicture}
        \begin{axis}[
            width=7.5cm,height=6.5cm,
            xmin=0, xmax=30,
            ymin=0, ymax=20,
            ytick={0,5,10,15,20},
            xlabel={\small $E_b/N_0$ (dB)},
            ylabel={\small Sum-Rate (bits/channel-use)},
            grid=both,
            grid style=dotted,
%              legend style={
%     at={(1.05,0.5)}, % right side
%     anchor=west,
%     font=\scriptsize
% },
%             legend cell align=left,
%             % legend entries={MC-RSMA with variant A, MC-RSMA with variant B, SCMA, Exact RS-SCMA, Jensen Approx. RS-SCMA},
%              legend entries={MC-RSMA , SCMA, Exact RS-SCMA, Jensen Approx. RS-SCMA},
%             cycle list name=color list,
legend style={
    at={(0.01,0.99)}, 
    anchor=north west
},
legend cell align=left,
            legend entries={\footnotesize MC-RSMA , \footnotesize SCMA,\footnotesize Exact RS-SCMA, \footnotesize RS-SCMA (LB)},
            cycle list name=color list,
        ]
            \addplot [green,thick,mark=triangle,mark size=3pt] 
            table [x={x}, y={y_var_b}, col sep=space]{sr_mc_rsma.txt};
            % \addplot [cyan,thick,mark=triangle,mark size=3pt] 
            % table [x={x}, y={y_var_a}, col sep=space]{sr_mc_rsma.txt};
            % \addplot  [red,thick,mark=square,mark size=3pt]  
            % table [x={x}, y={scma}, col sep=space]{scma_sr.txt};
             \addplot  [red,thick,mark=square,mark size=3pt]  
            table [x={x}, y={scma}, col sep=space]{sumrate_comp1.txt};
            % \addplot [black,thick,mark=oplus,thick,mark size=3pt, mark options={scale=1,solid}] 
            % table [x={x}, y={y_exact}, col sep=space]{sumrate_rsscma_new.txt};
            % \addplot [black,thick,mark=oplus*,dashed,mark size=3pt, mark options={scale=1,solid}] 
            % table [x={x}, y={y_lb}, col sep=space]{sumrate_rsscma_new.txt};
             \addplot [black,thick,mark=oplus,thick,mark size=3pt, mark options={scale=1,solid}] 
            table [x={x}, y={y_exact}, col sep=space]{rsscma_sr.txt};
            \addplot [black,thick,mark=oplus*,dashed,mark size=3pt, mark options={scale=1,solid}] 
            table [x={x}, y={y_jen}, col sep=space]{rsscma_sr.txt};
        \end{axis}
    \end{tikzpicture}
    \caption{\footnotesize{\color{black} Comparison of sum-rate for different SISO systems with $J=6$ users and $K=4$ subcarriers.}}
    \label{fig:sum_rate_r1}
\end{figure}

To provide a complete picture of the tunable system, Fig.~\ref{fig:sum_rate2} illustrates the effect of the splitting factor $\alpha$ on the sum-rate performance across its full range. These curves are generated using the unified analytical model from \eqref{eq:unified_final_rate}, which provides a fair comparison between the different operating points.
The plot clearly shows that the highest sum-rate is achieved at $\alpha = 0.5$. This curve serves as the performance benchmark, representing the system operating at its most spectrally efficient point where only Phase 1 (joint transmission) takes place. As $\alpha$ is tuned away from this central point, the sum-rate declines. When $\alpha$ decreases from $0.5$, the system spends more time in the private-only SCMA phase, with the $\alpha=0$ curve correctly converging to the performance of a conventional SCMA system. Conversely, as $\alpha$ increases beyond $0.5$, the common-only phase becomes more dominant. This demonstrates that while the hybrid configurations outperform the boundary cases ($\alpha=0$ and $\alpha=1$), the balanced split of $\alpha=0.5$ yields the maximum throughput.
\begin{figure}[!htpb]
    \centering
    \begin{tikzpicture}
        \begin{axis}[
            width=0.7\columnwidth,
            height=0.6\columnwidth,
            xmin=0, xmax=30,
            ymin=2, ymax=20,
            ytick={0,4,8,12,16,20},
            xlabel={\small $E_b/N_0$ (dB)},
            ylabel={\small Sum-Rate (bits/channel-use)},
            grid=both,
            grid style=dotted,
            legend style={
    at={(1.05,0.5)}, % right side
    anchor=west,
    font=\scriptsize
},
            legend cell align=left,
            legend entries={$\alpha=1$,$\alpha=0.75$,$\alpha=0.6$,
                            $\alpha=0.5$,$\alpha=0.4$,$\alpha=0.25$,$\alpha=0$},
        ]
            \addplot [thick,mark=triangle*,mark size=2pt,color=blue] 
            table [x={x}, y={y1}, col sep=space]{rsscma_alpha_new.txt};
            \addplot [thick,mark=square*,mark size=2pt,color=red]  
            table [x={x}, y={y75}, col sep=space]{rsscma_alpha_new.txt};
             \addplot [thick,mark=pentagon,mark size=2pt,color=pink!60!black] 
            table [x={x}, y={y60}, col sep=space]{rsscma_alpha_new.txt};
            \addplot [thick,mark=diamond*,mark size=2pt,color=green!60!black] 
            table [x={x}, y={y50}, col sep=space]{rsscma_alpha_new.txt};
             \addplot [thick,mark=pentagon*,mark size=2pt,color=cyan!60!black] 
            table [x={x}, y={y40}, col sep=space]{rsscma_alpha_new.txt};
            \addplot [thick,mark=o,mark size=2pt,color=orange] 
            table [x={x}, y={y25}, col sep=space]{rsscma_alpha_new.txt};
            \addplot [thick,mark=*,mark size=2pt,color=purple] 
            table [x={x}, y={y0}, col sep=space]{rsscma_alpha_new.txt};

        \end{axis}
    \end{tikzpicture}
    \caption{\footnotesize\color{black} Comparison of RS-SCMA sum-rate for different values of splitting factor $\alpha$.}
    \label{fig:sum_rate2}
\end{figure}

Fig.~\ref{fig:SIC_epsilon1_ed} investigates the impact of imperfect SIC on the sum-rate performance of the RS-SCMA system, parameterized by the  SIC imperfection factor $\epsilon$ used in \eqref{eq:sys_rx_private_main}. The case of $\epsilon = 0$ represents perfect SIC, where the common stream is completely removed, yielding the highest achievable sum-rate as an upper bound. When SIC is imperfect ($\epsilon > 0$), a fraction of the common stream remains as residual interference, directly impacting the detection of the private streams and thus lowering the overall sum-rate. At low SNRs, the system is noise-limited, and the performance loss due to imperfect SIC is minimal. %However, as the SNR increases, the system transitions into an interference-limited regime. 
In this high-SNR region, the system transitions from being noise-limited to interference-limited. The residual power from imperfect SIC becomes the dominant impairment, causing the sum-rate curves for $\epsilon>0$ to flatten into a distinct interference floor. This ceiling on performance prevents the sum-rate from growing further, regardless of increases in transmit power. As expected, a larger imperfection factor $\epsilon$ leads to a more pronounced performance degradation and a lower sum-rate ceiling. This analysis highlights the critical importance of accurate common stream decoding and cancellation for realizing the full potential of the RS-SCMA framework in practical, high-SNR scenarios.

\begin{figure}[!htpb]
    \centering
    \begin{tikzpicture}
        \begin{axis}[
            width=0.7\columnwidth,
            height=0.6\columnwidth,
            xmin=0, xmax=32,
            ymin=0, ymax=21,
            xtick={0,8,16,24,32},
            ytick={0,5,10,15,20},
            xlabel={\small $E_b/N_0$ (dB)},
            ylabel={\small Sum-Rate (bits/channel-use)},
            grid=both,
            grid style=dotted,
            legend style={
                at={(1.05,0.5)}, % right side
                anchor=west,
                font=\scriptsize
            },
            legend cell align=left,
            legend entries={\footnotesize $\epsilon=0$,
            \footnotesize $\epsilon=0.1$,
                            \footnotesize $\epsilon=0.3$,
                            \footnotesize $\epsilon=0.5$,
                            \footnotesize $\epsilon=0.7$,
                            \footnotesize $\epsilon=0.9$,
                            },
        ]
            % Plot for epsilon = 0
            \addplot [thick,blue,mark=triangle*,mark size=2.5pt] 
            table [x={x}, y={x0}, col sep=space]{eps_new.txt};
            \addplot [thick,red,mark=diamond,mark size=2.5pt] 
            table [x={x}, y={y1}, col sep=space]{eps1.txt};
            % Plot for epsilon = 0.2
            % \addplot [thick,red,mark=square*,mark size=2.5pt]  
            % table [x={x}, y={x1}, col sep=space]{eps_new.txt};

            % Plot for epsilon = 0.2
            \addplot [thick,green!60!black,mark=diamond*,mark size=2.5pt] 
            table [x={x}, y={x2}, col sep=space]{eps_new.txt};
            % \addplot [thick,magenta!60!black,mark=diamond,mark size=2.5pt] 
            % table [x={x}, y={x4}, col sep=space]{eps_new.txt};
 \addplot [thick,magenta!20!black,mark=square*,mark size=2.5pt] 
            table [x={x}, y={x5}, col sep=space]{eps_new.txt};
            % Plot for epsilon = 0.4
            \addplot [thick,orange,mark=o,mark size=2.5pt] 
            table [x={x}, y={x7}, col sep=space]{eps_new.txt};

            % Plot for epsilon = 0.5
            \addplot [thick,purple,mark=star,mark size=2.5pt] 
            table [x={x}, y={x9}, col sep=space]{eps_new.txt};
        \end{axis}
    \end{tikzpicture}
    \caption{\footnotesize Comparison of sum-rate for RS-SCMA for different values of SIC impact factor $\epsilon$.}
    \label{fig:SIC_epsilon1_ed}
\end{figure}

\subsection{Error-Rate Analysis}\label{ber}
Fig.~\ref{com_less} and~\ref{priv_less} show the BER performance of the RS-SCMA system for different values of the split factor \(\alpha\), which controls the distribution between common and private message symbols. As defined in~\eqref{lclp}, reducing \(\alpha\) increases \(l_p\), the number of private symbols, making the system operate more like conventional SCMA. In this case, private symbols benefit from multidimensional codewords and detection using the MPA, resulting in better interference handling and improved BER performance.
%For comparison, we consider the MISO-RSMA system as well. In the MISO-RSMA configuration, each user transmits two bit streams. By equally splitting these streams, a common message of 2 bits is formed, along with two private streams, each carrying 1 bit. In the work {\newcite{pam_rsma}}, it is shown that PAM has superior error rate performance than QAM, so, 4-PAM modulation is applied to the common message, while BPSK is used to modulate the private messages for all users.

Fig.~\ref{com_less} shows the BER performance for the private-dominant regime ($\alpha \le 0.5$), where a larger portion of the user message is allocated to the private SCMA layer. The results clearly demonstrate that reliability improves as the system moves away from the point of maximum overloading. At $\alpha=0.5$, the system operates at its peak overloading factor of 250\%, resulting in the highest BER. In this highly congested state, the common stream creates significant interference that must be canceled before the private messages can be decoded. As $\alpha$ is reduced (e.g., to 0.25 and 0.1), two key effects combine to significantly enhance BER performance. First, the system becomes less overloaded, and the interfering common stream decreases. This reduces the burden on the SIC stage. Second, and more importantly, a larger fraction of the user data is now encoded into the robust, multi-dimensional SCMA codewords. The powerful MPA detector is specifically designed to resolve MUI in the code domain. By shifting the system's load towards this more efficient MPA-based detection, the overall BER improves. This highlights that in the private-dominant regime, reducing the common stream's footprint directly enhances the effectiveness of the SCMA component, leading to superior error performance.

\pgfplotsset{every semilogy axis/.append style={
				line width=1 pt, tick style={line width=0.7pt}}, width=7.5cm,height=6cm, tick label style={font=\small},
			label style={font=\small},
			legend style={font=\scriptsize},
			legend pos= south west
            }           % Initial position

			\begin{figure}[!htpb]
				\centering
				\begin{tikzpicture}[new spy style]
					\begin{semilogyaxis}[xmin=0, xmax=30, ymin=4e-5, ymax=0.1,
						xlabel={\small $E_b/N_0$ (dB)},
                        xtick={0,5,10,15,20,25,30}, 
						ylabel={\small BER},
						grid=both,
						grid style={dotted}, % Bottom left corner
    legend cell align=left,
    font=\footnotesize, 
						legend entries={
    {$\footnotesize \alpha=0.1,\ \lambda_{\textnormal{RS-SCMA}}=167\%$},
    {$\footnotesize \alpha=0.25,\ \lambda_{\textnormal{RS-SCMA}}=195\%$},
    {$\footnotesize \alpha=0.5,\ \lambda_{\textnormal{RS-SCMA}}=250\%$}
},
						 cycle list name=mycolorlist,
                          legend pos=south west, % Legend base position
                         legend style={
    font=\scriptsize, % reduce font size
    fill=white,
    fill opacity=0.8,
    draw=none,
    at={(-0.001,-0.001)}, % bottom-left corner with margin
    anchor=south west
}, % Move legend right and up
						]
						%\addplot [green,thick,mark=square,mark size=3pt] table [x={x}, y={y1}] {ser_lp_more.txt};
					%	\addplot [black,thick,mark=oplus,thick,mark size=3pt, mark options={scale=1,solid}] table [x={x}, y={y2}] {ser_lp_more.txt};
						%\addplot [blue,thick,mark=diamond,mark size=3pt] table [x={x}, y={y}] {rsscma_soft250.txt};		
						\addplot [green,thick,mark=square*,mark size=3pt] table [x={x}, y={y10}] {ber_alpha_less.txt};
						\addplot [black,thick,mark=oplus*,thick,mark size=3pt, mark options={scale=1,solid}] table [x={x}, y={y25}] {ber_alpha_less.txt};
						\addplot [blue,thick,mark=diamond*,mark size=3pt] table [x={x}, y={y50}] {ber_alpha_less.txt};		
						
						%\addplot [green,thick,mark=triangle,mark size=3pt] table [x={x}, y={y}] {RSMA_uncoded.txt};
					\end{semilogyaxis}
				\end{tikzpicture}		
				\vspace*{-0.1cm}
				\caption{ \footnotesize{BER performance for RS-SCMA system with variable lengths of common and private message $(\alpha\leq 0.5,\ l_c\leq l_p)$  for $J=6,\ K=4$.}}
				\label{com_less}
			\end{figure}

Fig.~\ref{priv_less} considers the common-stream-dominant regime ($\alpha>0.5$) by varying the rate-splitting factor $\alpha$. In the proposed RS-SCMA framework, increasing $\alpha$ allocates a larger fraction of information to the common $M$-QAM stream and shortens the joint/private multiplexing portion; consequently, the effective overloading increases from $\lambda_{\mathrm{RS\text{-}SCMA}}\!\approx\!135\%$ at $\alpha=0.6$ to $\lambda_{\mathrm{RS\text{-}SCMA}}\!\approx\!225\%$ at $\alpha=0.9$. 

Despite this substantial change in $\lambda_{\mathrm{RS\text{-}SCMA}}$, the BER curves remain closely clustered over the practical SNR range (up to $\approx 30$~dB), indicating that the BER is only weakly sensitive to $\alpha$ in this operating region. This trend is consistent with the receiver structure: as $\alpha$ increases, a larger portion of the transmitted bits is conveyed by the common QAM layer, whose detection relies primarily on conventional  QAM demodulation and therefore cannot exploit the sparse multi-dimensional message-passing gain available to the SCMA private layers. Hence, tuning $\alpha$ mainly redistributes the payload between the common and private components, without yielding a proportional BER improvement. A slight separation is observed only at very high SNRs (above $\sim 30$~dB), where the lower-overloading configuration provides a marginal advantage. Moreover, pushing $\alpha$ further towards unity (common-only transmission) is not expected to significantly improve BER, since the dominant detection remains QAM-based rather than MPA-assisted.

            	\begin{figure}[!htpb]
    \centering
    \begin{tikzpicture}[new spy style]
        \begin{semilogyaxis}[
            xmin=0, xmax=35, ymin=1e-5, ymax=0,
            xlabel={\small $E_b/N_0$ (dB)},
            ylabel={\small BER}, xtick={0,5,10,15,20,25,30,35}, 
            grid=both,
            grid style={dotted},
            legend cell align=left,
            font=\footnotesize,
           legend entries={
    {$\tiny \alpha=0.9,\  \lambda_{\textnormal{RS-SCMA}}=135\%$},
    {$\tiny \alpha=0.75,\ \lambda_{\textnormal{RS-SCMA}}=183\%$},
    {$\tiny \alpha=0.6,\ \lambda_{\textnormal{RS-SCMA}}=225\%$}
},
            cycle list name=mycolorlist,
            legend pos=north east, % Legend base position
             legend style={
    font=\scriptsize, % reduce font size
    fill=white,
    fill opacity=0.8,
    draw=none,
    at={(-0.001,-0.001)}, % bottom-left corner with margin
    anchor=south west
}, % Move legend right and up
        ]
           % \addplot [black,thick,mark=oplus,thick,mark size=3pt, mark options={scale=1,solid}] table [x={x}, y={y90}] {lc_more1.txt};
            %\addplot [red,thick,mark=square,mark size=3pt] table [x={x}, y={y75}] {lc_more1.txt};
            %\addplot [blue,thick,mark=diamond,mark size=3pt] table [x={x}, y={y60}] {lc_more1.txt};		
            \addplot [black,thick,mark=oplus,thick,mark size=3pt, mark options={scale=1,solid}] table [x={x}, y={y90}] {ber_alpha_more.txt};
            \addplot [red,thick,mark=square,mark size=3pt] table [x={x}, y={y75}] {ber_alpha_more.txt};
            \addplot [blue,thick,mark=diamond,mark size=3pt] table [x={x}, y={y60}] {ber_alpha_more.txt};	
        \end{semilogyaxis}
    \end{tikzpicture}		
    %\vspace*{-0.1cm}
    \caption{\footnotesize{BER performance for RS-SCMA system with variable lengths of common and private message  $(\alpha> 0.5,\ l_c>l_p)$ for $J=6,\ K=4$.}}
    \label{priv_less}
\end{figure}
%%%%%%%%%%%%%%%%%%%%%%%%%%%%%%%%%%%%%%%%%%%%%%%%%%%%%%%%%%%%%%%%%
\pgfplotsset{
  every semilogy axis/.append style={
    line width=1pt,
    tick style={line width=0.7pt}
  },
  width=7cm,height=6cm,
  tick label style={font=\small},
  label style={font=\small},
  legend style={font=\small},
}
\begin{figure}[!htpb]
  \centering
  \begin{tikzpicture}
    \begin{semilogyaxis}[
      xmin=0, xmax=30, ymin=5e-5, ymax=1,
      xlabel={\small $E_b/N_0$ (dB)},
      ylabel={\small BER},
      grid=both,
      grid style={dotted},
      legend style={at={(0.01,0.18)}, anchor=west},
      legend cell align=left,
      legend pos=south west,
      legend entries={
        {\tiny SCMA ($\lambda=150\%$)},
        {\tiny SISO MC-RSMA},%{\tiny MISO $2\times 1$ MC-RSMA},
         %{\tiny MISO $4\times 1$ MC-RSMA},
        {\tiny RS\text{-}SCMA ($\lambda=250\%$)},
        {\tiny SCMA ($\lambda=250\%$)}
      },
      % remove the next line unless you've defined this cycle list
      % cycle list name=mycolorlist,
    ]
      \addplot [blue,thick,mark=diamond,mark size=3pt] table [x={x}, y={y_scma}] {scma1.txt};
      \addplot [red,thick,mark=square,mark size=3pt] table [x={x}, y={y1}] {mc_rsma_ber.txt};
       %\addplot [cyan,thick,mark=square*,mark size=3pt] table [x={x}, y={y2}] {mc_rsma_ber.txt};
       % \addplot [magenta,thick,mark=oplus,mark size=3pt] table [x={x}, y={y4}] {mc_rsma_ber.txt};
      \addplot [black,thick,mark=diamond*,mark size=3pt] table [x={x}, y={y50}] {ber_alpha_less.txt};
      \addplot [green,thick,mark=triangle,mark size=3pt] table [x={x}, y={y250}] {scma250.txt};
    \end{semilogyaxis}
  \end{tikzpicture}
  \vspace*{-0.1cm}
  \caption{\footnotesize BER performance comparison of the proposed SISO RS-SCMA with conventional SISO-SCMA and SISO MC-RSMA systems.}
  \label{diff_overload}
\end{figure}

{\color{black}
Fig.~\ref{diff_overload} evaluates the BER performance of the proposed RS-SCMA framework against the benchmarks. The conventional SCMA system with $J=6$ users ($\lambda_{\text{SCMA}}=150\%$, blue line) yields the most reliable detection, establishing a baseline for performance under moderate loading. 
A conventional SCMA with 250\% overloading factor with significantly larger number of users ($J=15$) is given in the plot, which drastically increases the density of the factor graph and the severity of multi-user interference. This severe contention overwhelms the MPA detector, resulting in the degraded performance observed in the green line. 

While RS-SCMA scheme (black line) increases the effective overloading factor to $\lambda_{\text{RS-SCMA}}=250\%$ yet maintains a significant performance advantage over the conventional SCMA configured for the same overloading (green line). Conversely, RS-SCMA achieves $\lambda_{\text{RS-SCMA}}=250\%$ while maintaining the sparser collision profile of $J=6$ users. By treating a portion of the interference as a decodable common message, the SIC stage effectively reduces the interference seen by the MPA detector. Crucially, in RS-SCMA, the MPA only needs to resolve contention among the original 6 users, not 15 and still the system has an overloading factor of 250\%. This allows it to operate far more effectively, demonstrating the architectural benefit of the hybrid approach.

Finally, the SISO MC-RSMA benchmark (red line) exhibits the highest error rate. Unlike the SCMA-based schemes, which benefit from multidimensional constellation shaping and spreading diversity, SISO MC-RSMA relies solely on power differences and SIC for user separation. Lacking both the spatial degrees of freedom (beamforming) and the structured code-domain sparsity, MC-RSMA fails to manage the multi-user interference effectively in this single-antenna setup.}\\
\textbf{Remark 3:} \textit{QAM symbols are complex scalars, whereas SCMA maps information onto multidimensional complex vector codewords. Despite operating under user overloading, SCMA achieves a steeper BER slope compared to conventional $M$-QAM, owing to its sparse codeword structure, coding and shaping gain and the use of MPA detection.}\\
%%%%%%%%%%%%%%%%%%%%%%%%%%%%%%%%%%%%%%%%%%%%
\begin{figure}[!t]
\centering
\begin{tikzpicture}

\begin{groupplot}[
  group style={group size=1 by 2, vertical sep=14pt}, % enough space for alpha
  width=0.8\columnwidth,
  height=0.55\columnwidth,
  xmin=0, xmax=1,
  xtick={0,0.2,0.4,0.6,0.8,1},
  grid=major,
  grid style={dotted,gray!35},
  tick align=outside,
  tick style={black},
  tick label style={font=\footnotesize},
  axis line style={line width=0.8pt},
  legend to name=combinedlegend,
  legend style={
    draw=none, fill=white, fill opacity=0.9,
    font=\footnotesize, legend columns=3,
    /tikz/every even column/.append style={column sep=6pt}
  }
]

% ---------- (a) Overloading factor ----------
\nextgroupplot[
  axis lines=box,
  tick pos=left,
  xticklabels={}, % remove x tick labels in top plot
  %xtick=\empty,   % remove x ticks entirely
  xlabel=\empty,
  ylabel={\footnotesize \textcolor{red}{$\lambda_{\text{RS-SCMA}}$}},
  ymin=100, ymax=260
]

%\addplot [forget plot, red, line width=1.0pt, mark=o, mark size=2.5pt]
\addplot [forget plot, red, solid, line width=1.0pt, no marks]
%\addlegendimage{red, solid, line width=1.0pt, no marks}
  table [x={x}, y={y}, col sep=space]{lc_vs_lambda.txt};

% ---------- (b) Sum-rate (bottom, left axis) ----------
\nextgroupplot[
  axis lines=left,
  ylabel={\footnotesize \textcolor{blue}{Sum-rate (bits/channel-use)}},
  yticklabel style={text=blue},
  y label style={text=blue},
  ymin=8, ymax=20,
  %xticklabel pos=top % move bottom plot’s ticks to the top
  xlabel={\footnotesize Splitting factor $\alpha$},
]
\addplot [forget plot, blue, dashed, line width=1.0pt, mark=triangle*, mark size=2.5pt]
  table [x={x}, y={SR}, col sep=space]{srvsalpha_new.txt};

% ---- Unified legend entries (manual, fixed order) ----
% \addlegendimage{red, solid, line width=1.0pt, mark=o, mark size=2.5pt}
\addlegendimage{red, solid, line width=1.0pt, no marks}
\addlegendentry{Overloading Factor}
\addlegendimage{blue, dashed, line width=1.0pt, mark=triangle*, mark size=2.5pt}
\addlegendentry{Sum-rate at 30 dB}
\addlegendimage{green!40!black, solid, line width=1.0pt, mark=square*, mark size=2.5pt}
\addlegendentry{BER at 30 dB}
\addlegendimage{black, dotted, line width=1.0pt}
\addlegendentry{Lower Bound for BER}

\end{groupplot}
% Shared x-axis label between the two plots
% \node[font=\footnotesize, anchor=center]
%   at ($(group c1r1.south)!0.5!(group c1r2.north)$) {Splitting factor $\alpha$};

% ---------- (c) BER overlay (bottom, right axis) ----------
\begin{axis}[
  width=0.8\columnwidth,
  height=0.55\columnwidth,
  at={(group c1r2.south west)}, anchor=south west,
  xmin=0, xmax=1,
  ymode=log,
  ymin=5e-6, ymax=2e-3,
  axis y line*=right,
  axis x line=none,
  xmajorticks=false,
  ylabel={\footnotesize \textcolor{green!40!black}{BER}},
  yticklabel style={font=\footnotesize}
]
\addplot [forget plot, green!40!black, solid, line width=1.0pt, mark=square*, mark size=2.5pt]
  table [x={alpha}, y={ber30}, col sep=space]{bervsalpha.txt};
  \addplot [forget plot, black, dotted, line width=1.0pt]
  table [x={alpha}, y={LB}, col sep=space]{bervsalpha.txt};
\end{axis}

% ---------- Alpha label between the plots ----------
% \node[font=\footnotesize, anchor=center]
%   at ($(group c1r1.south)!0.5!(group c1r2.north)$) {Splitting factor $\alpha$};

\end{tikzpicture}

% Place the legend below both plots
\pgfplotslegendfromname{combinedlegend}

\caption{\footnotesize RS\text{-}SCMA: Overloading factor (top) and sum-rate (left) with BER (right, log scale) vs.\ splitting factor $\alpha$.}
\label{fig:combined_plot}
\end{figure}
Fig.~\ref{fig:combined_plot} provides a comprehensive analysis of the system's key performance metrics as a function of the splitting factor $\alpha$, evaluated at a fixed $E_b/N_0 = 30\,\mathrm{dB}$. This analysis clearly illustrates the fundamental trade-offs inherent in the RS-SCMA framework. The top panel shows that the overloading factor, $\lambda_{\text{RS-SCMA}}$, peaks at 250\% at the balanced point of $\alpha=0.5$. The bottom panel reveals the direct consequences of this on throughput and reliability. The sum-rate (blue dashed line) mirrors the overloading trend, achieving a maximum of approximately 20~bits/channel-use at $\alpha=0.5$. This confirms that the configuration gets maximum throughput and is most spectrally efficient when only joint transmission prevails.

However, a clear rate-reliability trade-off is evident in the BER curve (solid green line). The best BER performance (lowest error rate) is achieved at $\alpha=0$, which corresponds to a conventional SCMA system. As common symbols are introduced (increasing $\alpha$ towards 0.5), the system becomes more loaded and the mutual interference between layers increases, causing the BER to degrade, reaching its worst point near $\alpha=0.5$. To benchmark this performance, a theoretical lower bound for the BER is also plotted (dotted line) which represents an idealized interference-free system and is calculated as a weighted average of the BERs of the standalone common-only transmission ($\alpha=1$) and private-only ($\alpha=0$ SCMA) systems:
\setcounter{equation}{44}
\begin{equation}
\label{eq:ber_lower_bound}
\text{BER}_{\text{LB}}(\alpha) = \alpha  \text{BER}_{\text{common-only}} + (1-\alpha)  \text{BER}_{\text{SCMA}}.
\end{equation}Interestingly, in the common-dominant regime ($\alpha > 0.5$), the BER improves from its worst point and remains relatively stable, closely approaching this theoretical lower bound. This demonstrates that while the joint transmission is active, the powerful SCMA based MPA detector for the private streams effectively manages the multi-user interference, maintaining a robust error performance even at high overloading factors. This analysis highlights the crucial flexibility of the RS-SCMA framework. The splitting factor $\alpha$ serves as a practical tuning to navigate the trade-off between maximizing system capacity and ensuring user reliability. While pure SCMA ($\alpha=0$) offers the highest reliability, a slight shift to $\alpha=0.25$ provides a massive boost in sum-rate with only a minor penalty in BER. Conversely, choosing $\alpha=0.5$ delivers the absolute maximum throughput for applications where a higher error rate is tolerable.

                        %%%%%%%%%%%%%%%%LDPC%%%%%%%%%%%%%%%%%%%%%%%%

 \pgfplotsset{every semilogy axis/.append style={
				line width=1 pt, tick style={line width=0.7pt}}, width=7.5cm,height=6cm, tick label style={font=\small},
			label style={font=\small},
			legend style={font=\small},
			legend style={at={(0.99,0.99)}, anchor=north east}}
\begin{figure}[!htpb]
	\centering
	\begin{tikzpicture}[new spy style]
		\begin{semilogyaxis}[xmin=8, xmax=20, ymin=2e-5, ymax=0,
			xlabel={\small $E_b/N_0$ (dB)},
			ylabel={\small BLER},
			grid=both,
			grid style={dotted},
			legend cell align=left,
			legend entries={\footnotesize LDPC-Coded Rx-1,\footnotesize Rx-2,\footnotesize SCMA($\lambda=250\%$),\footnotesize MC-RSMA },
			cycle list name=mycolorlist,
			]
			
			\addplot [blue,thick,mark=oplus,mark size=3pt] table [x={x}, y={y2}] {fading_rec.txt};	
			\addplot [black,thick,mark=triangle,mark size=3pt] table [x={x}, y={y1}] {fading_rec.txt};	
            \addplot [green,thick,mark=diamond,mark size=3pt] table [x={x1}, y={y4}] {ldpc_comp1_1.txt};	%%scma250
						\addplot [red,thick,mark=square,mark size=3pt] table [x={x}, y={y_rsma}] {mc_rsma_ldpc.txt};	
		\end{semilogyaxis}
	\end{tikzpicture}		
	\caption{\footnotesize{BLER performance comparison for LDPC-coded Rx-1, Rx-2 with LDPC coded SISO-SCMA and coded SISO MC-RSMA.}}
	\label{arch_ldpc_1}
\end{figure}
{\color{black}Fig.~\ref{arch_ldpc_1} evaluates the BLER performance of the LDPC-coded systems. We configure the proposed RS-SCMA framework for its peak throughput mode ($\alpha = 0.5$, $\lambda_{\text{RS-SCMA}} = 250\%$) and set the SCMA benchmark to a matching overloading factor ($\lambda_{\text{SCMA}} = 250\%$). 
To ensure a rigorous comparison, we carefully select the code rates from the 5G-NR LDPC family. For the proposed RS-SCMA, we adopt a balanced configuration of $r_c = r_p = 0.468$. In contrast, for the SCMA and MC-RSMA benchmarks, we assign a significantly stronger code with a lower rate of $r = 0.323$. This configuration provides the baselines with enhanced error-correction capability, ensuring that any performance gains achieved by RS-SCMA are attributable to its architectural efficiency rather than coding advantages. These coded scenarios serve to validate that the reliability trends observed in the uncoded analysis persist under practical channel coding constraints.

The results demonstrate that even with the stronger code, the SISO MC-RSMA benchmark suffers from high interference sensitivity. The heavily overloaded SCMA ($\lambda_{\text{SCMA}}=250\%$) outperforms MC-RSMA, confirming the robustness of sparse code-domain multiplexing. However, the proposed RS-SCMA framework yields the highest reliability among all configurations, despite utilizing a higher code rate ($0.468$ vs $0.323$). Specifically, the Rx-2 receiver utilizes refined soft information from the channel decoder to aid the interference cancellation process. This effectively minimizes residual errors, allowing Rx-2 to deliver a notable coding gain of 0.839~dB over the simpler Rx-1 architecture at a target BLER of $10^{-3}$. The coded results are intended to validate the reliability trends observed in the uncoded 
    comparisons under practical channel coding, and are not meant to introduce any 
    new contribution in channel coding.
    
    Fig.~\ref{arch_ldpc_3} shows the code-rate allocation for RS-SCMA. 
The lowest BLER is achieved when both streams are strongly protected \((r_c = r_p = 0.323,\) red curve).
An important insight comes from the asymmetric allocations. 
Providing stronger protection to the common stream \((r_c = 0.323,\, r_p = 0.468,\) cyan curve) consistently outperforms the opposite case \((r_c = 0.468,\, r_p = 0.323,\) blue curve).
This performance gap cannot be explained only by the standard SIC behavior observed in SISO MC-RSMA, where reliability mainly depends on power allocation and decoding order. 
In RS-SCMA, the two layers are physically different. 
The private streams are carried by sparse SCMA codewords and detected using the MPA, which offers intrinsic multidimensional shaping gain. 
In contrast, the common stream uses conventional QAM and lacks similar structural robustness.

Consequently, the common stream becomes the reliability bottleneck, so it must be decoded first. 
When the common stream is assigned a higher code rate (blue curve), decoding fails early. 
This causes SIC failure and makes the subsequent MPA stage ineffective. 
Therefore, the common stream should use a lower code rate. 
This compensates for the modulation scheme  mismatch and stabilizes the overall RS-SCMA detection chain.
}

\pgfplotsset{every semilogy axis/.append style={
				line width=1 pt, tick style={line width=0.7pt}}, width=7.5cm,height=6cm, tick label style={font=\small},
			label style={font=\small},
			legend style={font=\small},
			legend pos= north east}
\begin{figure}[!htpb]
    \centering
    \begin{tikzpicture}[new spy style]
        \begin{semilogyaxis}[xmin=8, xmax=16, ymin=9e-6, ymax=0.02,
            xlabel={\small $E_b/N_0$ (dB)},
            ylabel={\small BLER},
            grid=both,
            grid style={dotted},
            legend style={at={(1.03,1.03)},anchor=north east},
            font=\tiny, 
            legend entries={$r_c=0.323\ r_p=0.323$,
            $r_c=0.468\ r_p=0.468$ , 
            $r_c=0.468\ r_p=0.323$,$r_c=0.323\ r_p=0.468$ },
            cycle list name=mycolorlist,
            ]
           % \addplot [green,thick,mark=oplus,thick,mark size=3pt, mark options={scale=1,solid}] table [x={x}, y={y}] {rs_scma_ldpc_soft.txt};
            %\addplot [magenta,thick,mark=square,mark size=3pt] table [x={x}, y={y}] {rs_scma_ldpc3.txt};
           % \addplot [blue,thick,mark=triangle,mark size=3pt] table [x={x}, y={y}] {ser12.txt};
           % \addplot [blue,thick,mark=diamond,mark size=3pt] table [x={x}, y={y1}] {ldpc_diffrate.txt};
            %\addplot [blue,thick,mark=star,mark size=3pt] table [x={x}, y={y}] {rc13rp12.txt};
           % \addplot [red,thick,mark=diamond,mark size=3pt] table [x={x}, y={y}] {rc12rp13.txt};
           \addplot [red,thick,mark=diamond*,mark size=3pt] table [x={x}, y={y}] {rcrp13.txt};
           \addplot [black,thick,mark=triangle*,mark size=3pt] table [x={x}, y={y1}] {fading_rec.txt};
          % \addplot [blue,thick,mark=star,mark size=3pt] table [x={x}, y={y}] {rc23rp12_1.txt};
            \addplot [blue,thick,mark=diamond,mark size=3pt] table [x={x}, y={y}] {rc12rp13_new.txt};
          \addplot [cyan,thick,mark=star,mark size=3pt] table [x={x}, y={y}] {rc13rp12_fin3.txt};
         
        \end{semilogyaxis}
    \end{tikzpicture}       
    \caption{\footnotesize{BLER performance for different LDPC code rates for common and private messages in the RS-SCMA system  with $J=6$ and $K=4$ for Rx-2.}}
    \label{arch_ldpc_3}
\end{figure}

\begin{figure}[!htpb]
    \centering
    \begin{tikzpicture}[new spy style]
        \begin{semilogyaxis}[
            width=7.5cm,
            height=6cm,
            xmin=0, xmax=30, ymin=3e-4, ymax=0.2,
            xlabel={\scriptsize $E_b/N_0$ (dB)},
            ylabel={\scriptsize BER},
            grid=both,
            grid style={dotted},
            font=\scriptsize,
            legend style={
                at={(0.02,0.02)}, % bottom-left corner
                anchor=south west,
                font=\scriptsize
            },
            legend entries={$e=0\%$,$e=1\%$,$e=4\%$,$e=6\%$,$e=10\%$},
        ]

            % e = 0%

            \addplot [red,thick,mark=diamond*,mark size=2pt] 
                table [x={x}, y={y15}] {csir.txt};

            % e = 2%
             \addplot [blue,dotted,mark=square*,mark size=2pt] 
                table [x={x}, y={y10}] {csir.txt};
            % e = 4%
            \addplot [green!60!black,thick,mark=o,mark size=2pt] 
                table [x={x}, y={y7}] {csir.txt};

            % e = 6%
            \addplot [magenta,thick,mark=triangle*,mark size=2pt] 
                table [x={x}, y={y4}] {csir.txt};

            % e = 10%
            \addplot [black,thick,mark=star,mark size=2pt] 
                table [x={x}, y={y2}] {csir.txt};

        \end{semilogyaxis}
    \end{tikzpicture}       
    \caption{\footnotesize BER analysis of RS-SCMA under imperfect channel estimation for varying $e$ values.}
    \label{csir_ed}
\end{figure}

To analyse the impact of imperfect CSIR, we model the estimated channel $\hat{\mathbf{H}}_j$ for the \( j \)th user denoted as:
\begin{equation}
\hat{\mathbf{H}}_j = \mathbf{H}_j + \sqrt{e} \boldsymbol{\Omega}_j,
\label{imp_H_ed}
\end{equation}
where \( \mathbf{H}_j \in \mathbb{C}^{K \times K} \) is the actual diagonal channel matrix for user \( j \), and \( \boldsymbol{\Omega}_j \in \mathbb{C}^{K \times K} \) is a diagonal matrix whose entries are i.i.d. complex Gaussian random variables with zero mean and unit variance, i.e., \( \mathcal{CN}(0,1) \). Here, \( e \) denotes the variance of the estimation error. The matrices \( \hat{\mathbf{H}}_j \) and \( \boldsymbol{\Omega}_j \) are assumed to be uncorrelated.
Fig.~\ref{csir_ed} illustrates the BER performance of the RS-SCMA system under imperfect CSIR, modeled by~\eqref{imp_H_ed}. As the error variance $e$ increases from 0 (perfect CSIR) to 0.01 (representing a 10\% error standard deviation), the BER degrades across all $E_b/N_0$ values. The BER degrades across all \(E_b/N_0\) values, with the degradation becoming more pronounced at high SNR. This occurs because at high SNR, the noise level is low, and even small inaccuracies in the estimated channel \(\hat{\mathbf{H}}_j\) lead to mismatches in the SIC. These mismatches generate residual interference that cannot be effectively cancelled, especially in the decoding of private symbols using the MPA. As a result, error propagation occurs and leads to a BER floor, highlighting the sensitivity of RS-SCMA to channel estimation errors in high-SNR regimes.

\section{Conclusions and Future Works}\label{conc}
{\color{black}
% This paper introduced SISO RS-SCMA, the first framework to  integrate RS into a CD-NOMA architecture. In this hybrid design for downlink communication, a tunable splitting factor is introduced to govern the allocation between $M$-QAM modulated common messages and SCMA-encoded private messages, providing direct control over the fundamental trade-off between the system capacity and user reliability. Crucially, unlike SISO MC-RSMA, which requires CSIT, our framework relies on code-domain multiplexing and receiver-side processing, making it a practical solution for massive machine-type communication networks. We developed novel transmitter and receiver architectures and derived a unified analytical expression for the achievable sum-rate that accounts for practical imperfections such as SIC errors and imperfect channel state information. Our extensive simulation results validate this theoretical framework, demonstrating that SISO RS-SCMA consistently outperforms conventional SISO SCMA and SISO MC-RSMA in terms of BER, BLER, and sum-rate, even under channel estimation errors. With the tunable splitting factor $\alpha$ introduced to allocate information bits to the common and private layers. The analysis confirms that the splitting factor is an effective tool for navigating the rate-reliability trade-off. Future work will explore the joint optimization of this factor with dynamic power allocation, as well as extensions to MIMO systems with precoder design and imperfect CSIT.

This paper introduced a SISO RS-SCMA framework that integrates rate-splitting into a CD-NOMA architecture. For the downlink, we proposed a hybrid transmission design in which a tunable splitting factor $\alpha$ controls the allocation of information bits between an $M$-QAM-modulated common layer and SCMA-encoded private layers, thereby enabling direct control of the fundamental rate--reliability trade-off. Unlike MC-RSMA, which typically relies on CSIT-enabled precoding, the proposed framework is primarily driven by code-domain multiplexing and receiver-side processing, making it well-suited for practical massive machine-type communications.
We developed the corresponding transmitter and receiver architectures and derived a unified achievable sum-rate expression that explicitly incorporates practical impairments, including imperfect SIC and channel estimation errors. The simulations validate the analysis and demonstrate consistent gains over conventional SISO SCMA and SISO MC-RSMA in terms of BER, BLER, and sum-rate, including under imperfect CSIR. The results further confirm that $\alpha$ serves as an effective design knob for navigating the throughput--reliability trade-off. Future work will investigate joint optimization of $\alpha$ with dynamic power allocation and extend the framework to MIMO RS-SCMA with precoder design under imperfect CSIT.
}
%%%%%%%%%%%%%%%%%%%%%%%%%%%%%%%%%%%%%%%%%%% Appendix %%%%%%%%%%%%%%%
{\color{black}
\appendices 
%\section{Derivation of Finite-Alphabet Exact Rates and Jensen Lower Bound}

%\begin{appendices}
\section{Derivation of Finite-Alphabet Constrained Exact Rates and Jensen Lower Bound}
\label{app:jensen_fa}

In this appendix, we outline the key derivation for
(i) the exact finite-alphabet expressions in \eqref{eq:Rc_exact_main} and \eqref{eq:Rp_exact_main}, and
(ii) the approximated tractable lower bounds in \eqref{eq:Rc_LB_main} and \eqref{eq:Rp_LB_main}.
The derivation follows standard finite-alphabet information-theoretic techniques \cite{Wu,zeng_low,rsma_receiever}.

%\subsection{Generic Discrete-Input Continuous-Output Model}

Consider the $K$-dimensional model
\begin{equation}
\mathbf{y} = \mathbf{x} + \boldsymbol{\zeta} + \mathbf{n},
\qquad
\mathbf{n}\sim\mathcal{CN}(\mathbf{0},N_0\mathbf{I}_K),
\label{eq:app_model_generic}
\end{equation}
where $\mathbf{x}\in\mathcal{X}\subset\mathbb{C}^{K\times 1}$ is equiprobable over a finite alphabet $\mathcal{X}$ with $|\mathcal{X}|=M$.
The term $\boldsymbol{\zeta}$ denotes a discrete random vector induced by finite-alphabet interference; i.e.,
$\boldsymbol{\zeta}$ takes values in a discrete set $\mathcal{Z}\subset\mathbb{C}^{K\times 1}$ (the set of all possible interference realizations),
with PMF $P(\boldsymbol{\zeta})$.
We assume $\mathbf{x}$ is independent of $\boldsymbol{\zeta}$ and $\mathbf{n}$.

%\subsection{Exact Conditional Entropy: Log-Sum-Exp Form}

The achievable rate with discrete interference is
\begin{equation}
R^{\mathrm{exact}} = I(\mathbf{x};\mathbf{y})
= H(\mathbf{x}) - H(\mathbf{x}\mid \mathbf{y}),
\label{eq:app_MI_def}
\end{equation}
where $H(\mathbf{x})=\log_2 M$ for equiprobable symbols.
Using the law of total expectation over $\boldsymbol{\zeta}$, we get
\begin{equation}
H(\mathbf{x}\mid\mathbf{y})
=
\mathbb{E}_{\boldsymbol{\zeta}}
\!\left[
H(\mathbf{x}\mid \mathbf{y},\boldsymbol{\zeta})
\right],
\label{eq:app_total_expectation}
\end{equation}
%$\mathbf{y}$ is random due to both the discrete interference realization
%$\boldsymbol{\zeta}$ and the continuous Gaussian noise $\mathbf{n}$.

For a fixed $\boldsymbol{\zeta}$ and a transmitted $\mathbf{x}_a\in\mathcal{X}$, we have
$\mathbf{y} = \mathbf{x}_a+\boldsymbol{\zeta}+\mathbf{n}$.
Applying Bayes' rule with uniform priors yields the standard log-sum-exp form
% \begin{equation}
% H(\mathbf{x}\mid \mathbf{y},\boldsymbol{\zeta})
% =
% \frac{1}{M}\sum_{\mathbf{x}_a\in\mathcal{X}}
% \mathbb{E}_{\mathbf{n}}
% \left[
% \log_2
% \sum_{\mathbf{x}_b\in\mathcal{X}}
% \exp\!\left(
% -\frac{
% \left\|(\mathbf{x}_a-\mathbf{x}_b)+\boldsymbol{\zeta}+\mathbf{n}\right\|^2
% -
% \left\|\boldsymbol{\zeta}+\mathbf{n}\right\|^2
% }{N_0}
% \right)
% \right].
% \label{eq:app_logsumexp_generic}
% \end{equation}

\begin{equation}
\label{eq:app_logsumexp_generic}
\begin{aligned}
H(\mathbf{x}\mid \mathbf{y},\boldsymbol{\zeta})
&=
\frac{1}{M}\sum_{\mathbf{x}_a\in\mathcal{X}}
\mathbb{E}_{\mathbf{n}}
\Bigg[
\log_2
\sum_{\mathbf{x}_b\in\mathcal{X}}
\exp\!\left(
-\frac{\Delta_{ab}(\boldsymbol{\zeta},\mathbf{n})}{N_0}
\right)
\Bigg],
\end{aligned}
\end{equation}
where,
\begin{equation}
\label{eq:app_Delta_ab_generic}
\begin{aligned}
\Delta_{ab}(\boldsymbol{\zeta},\mathbf{n})
&\triangleq
\left\|(\mathbf{x}_a-\mathbf{x}_b)+\boldsymbol{\zeta}+\mathbf{n}\right\|^2
-
\left\|\boldsymbol{\zeta}+\mathbf{n}\right\|^2 .
\end{aligned}
\end{equation}

Combining \eqref{eq:app_total_expectation} and \eqref{eq:app_logsumexp_generic} with $H(\mathbf{x})=\log_2 M$ gives the exact sum-rate expression in \eqref{eq:app_MI_def}.

%\subsection{Specialization to RS-SCMA: Common and Private Layers}

\subsubsection*{A) Common-layer}
During common decoding at user $u$, set
\[
\mathbf{x}=\sqrt{p_c}\mathbf{H}_u\mathbf{s}_c,\quad
\mathcal{X}=\mathcal{S}_c,\quad
|\mathcal{S}_c|=M_c^K,\quad
\boldsymbol{\zeta}=\boldsymbol{\zeta}_{\mathrm{tot},u}\in\mathcal{Z}_{\mathrm{tot},u},
\]
where $\boldsymbol{\zeta}_{\mathrm{tot},u}$ and $\mathcal{Z}_{\mathrm{tot},u}$ are defined in
\eqref{eq:zeta_tot_def} and \eqref{eq:Ztot_def_main}, respectively.
Substituting into \eqref{eq:app_logsumexp_generic} yields \eqref{eq:Rc_exact_main} with the distance metric in \eqref{eq:Delta_c_main}.
Finally, that  gives $R_c^{\mathrm{exact}}=\min_u R_{c,u}^{\mathrm{exact}}$.

\subsubsection*{B) Private-layer }
After common decoding, the post-SIC model at user $u$ is \eqref{eq:sys_rx_private_main}.
Set
\[
\mathbf{x}=\sqrt{p_{p,u}}\mathbf{H}_u\mathbf{c}_u,\quad
\mathcal{X}=\mathcal{C}_u,\quad
M=M_p,\quad
\boldsymbol{\zeta}=\boldsymbol{\zeta}_{u}+\sqrt{\epsilon p_c}\mathbf{H}_u\mathbf{s}_{\mathrm{res}}.
\]
Here $\boldsymbol{\zeta}_{u}\in\mathcal{Z}_{u}$ with $\mathcal{Z}_u$ defined in \eqref{eq:Zu_def_main},
and $\mathbf{s}_{\mathrm{res}}\in\mathcal{S}_c$ is the residual common symbol.
Under independent equiprobable signaling,
$P(\boldsymbol{\zeta}_{u}) = 1/|\mathcal{Z}_u|$ and $P(\mathbf{s}_{c})=1/|\mathcal{S}_c|$, and thus
$P(\boldsymbol{\zeta}_{u},\mathbf{s}_{c})=P(\boldsymbol{\zeta}_{u})P(\mathbf{s}_{c})$.
Substituting into \eqref{eq:app_logsumexp_generic} yields \eqref{eq:Rp_exact_main} with \eqref{eq:Delta_p_main}.

%\subsection{Jensen Step and Closed-Form Gaussian Expectation}

The exact expressions require the expectation over $\mathbf{n}$ inside the $\log_2(\cdot)$ term.
To obtain a tractable bound, Jensen's inequality is applied only to the Gaussian-noise expectation:
\begin{equation}
\mathbb{E}_{\mathbf{n}}\!\left[\log_2 Z(\mathbf{n})\right]
\le
\log_2\!\left(\mathbb{E}_{\mathbf{n}}[Z(\mathbf{n})]\right),
\label{eq:app_jensen}
\end{equation}
which upper-bounds the conditional entropy and thus yields a lower bound on the mutual information.

The key Gaussian expectation takes the form
\begin{equation}
\mathbb{E}_{\mathbf{n}}
\left[
\exp\!\left(-\frac{\|\mathbf{v}+\mathbf{n}\|^2}{N_0}\right)
\right]
=
2^{-K}\exp\!\left(-\frac{\|\mathbf{v}\|^2}{2N_0}\right),
\label{eq:app_gauss_identity}
\end{equation}
which produces %(i) an effective variance doubling $N_0\to 2N_0$ inside the exponent and (ii) 
the constant gap term
\begin{equation}
\kappa \triangleq K\left(\frac{1}{\ln 2}-1\right).
\label{eq:app_kappa}
\end{equation}
Substituting \eqref{eq:app_gauss_identity} into \eqref{eq:app_jensen} and then into \eqref{eq:app_logsumexp_generic}
directly yields the tractable lower bounds in \eqref{eq:Rc_LB_main} and \eqref{eq:Rp_LB_main}.}

\bibliographystyle{IEEEtran}
			\footnotesize
            \bibliography{references_RSMA}

@ARTICLE{Wu,
  author={Wu, Yongpeng and Wang, Mingxi and Xiao, Chengshan and Ding, Zhi and Gao, Xiqi},
  journal={IEEE Transactions on Wireless Communications}, 
  title={{Linear Precoding for MIMO Broadcast Channels With Finite-Alphabet Constraints}}, 
  year={2012},
  volume={11},
  number={8},
  pages={2906-2920},
  doi={10.1109/TWC.2012.060412.111806}}

@ARTICLE{NOMA_survey,
  author={Dai, Linglong and Wang, Bichai and Yuan, Yifei and Han, Shuangfeng and Chih~lin, I. and Wang, Zhaocheng},
  journal={IEEE Commun. Mag.}, 
  title={{Non-orthogonal Multiple Access for {5G}: Solutions, Challenges, Opportunities, and Future Research Trends}}, 
  year={2015},
  volume={53},
  number={9},
  pages={74-81},
  doi={10.1109/MCOM.2015.7263349}}

@ARTICLE{NOMA_survey1,
  author={Ding, Zhiguo and Lei, Xianfu and Karagiannidis, George K. and Schober, Robert and Yuan, Jinhong and Bhargava, Vijay K.},
  journal={IEEE J. on Selected Areas in Commun.}, 
  title={{A Survey on Non-Orthogonal Multiple Access for {5G} Networks: Research Challenges and Future Trends}}, 
  year={2017},
  volume={35},
  number={10},
  pages={2181-2195},
  doi={10.1109/JSAC.2017.2725519}}

@ARTICLE{Chen2020,
  author={Chen, Hongzhi and Mi, De and Clerckx, Bruno and Chu, Zheng and Shi, Jia and Xiao, Pei},
  journal={IEEE Transactions on Vehicular Technology}, 
  title={{Joint Power and Subcarrier Allocation Optimization for Multigroup Multicast Systems With Rate Splitting}}, 
  year={2020},
  volume={69},
  number={2},
  pages={2306-2310},
  doi={10.1109/TVT.2019.2957994}}

@ARTICLE{Pereira2025,
  author={Pereira, Emanuel Valério and Lima, Francisco Rafael Marques},
  journal={IEEE Access}, 
  title={{Low-Complexity User Matching and Stream-Based Power Allocation for Multicarrier RSMA Systems}}, 
  year={2025},
  volume={13},
  number={},
  pages={191470-191484},
  doi={10.1109/ACCESS.2025.3630122}}

@ARTICLE{Chen2021,
  author={Chen, Hongzhi and Mi, De and Wang, Tong and Chu, Zheng and Xu, Yin and He, Dazhi and Xiao, Pei},
  journal={IEEE Transactions on Broadcasting}, 
  title={{Rate-Splitting for Multicarrier Multigroup Multicast: Precoder Design and Error Performance}}, 
  year={2021},
  volume={67},
  number={3},
  pages={619-630},
  doi={10.1109/TBC.2021.3071061}}

@ARTICLE{mc_rsma,
  author={Li, Lihua and Chai, Kejia and Li, Jilong and Li, Xingwang},
  journal={IEEE Access}, 
  title={Resource Allocation for Multicarrier Rate-Splitting Multiple Access System}, 
  year={2020},
  volume={8},
  number={},
  pages={174222-174232},
  doi={10.1109/ACCESS.2020.3025635}}

@ARTICLE{mmf_noma,
  author={Wang, Chin-Liang and Chen, Jyun-Yu and Chen, Yi-Jhen},
  journal={IEEE Wireless Communications Letters}, 
  title={{Power Allocation for a Downlink Non-Orthogonal Multiple Access System}}, 
  year={2016},
  volume={5},
  number={5},
  pages={532-535},
  doi={10.1109/LWC.2016.2598833}}

@ARTICLE{uma,
  author={Clerckx, Bruno and Mao, Yijie and Yang, Zhaohui and Chen, Mingzhe and Alkhateeb, Ahmed and Liu, Liang and Qiu, Min and Yuan, Jinhong and Wong, Vincent W. S. and Montojo, Juan},
  journal={Proceedings of the IEEE}, 
  title={{Multiple Access Techniques for Intelligent and Multifunctional 6G: Tutorial, Survey, and Outlook}}, 
  year={2024},
  volume={112},
  number={7},
  pages={832-879},
  doi={10.1109/JPROC.2024.3409428}}

@ARTICLE{capacity_based_SCMA,
  author={Xiao, Kexin and Xia, Bin and Chen, Zhiyong and Xiao, Baicen and Chen, Dageng and Ma, Shaodan},
  journal={IEEE Trans. on Wireless Commun.}, 
  title={{On Capacity-Based Codebook Design and Advanced Decoding for Sparse Code Multiple Access Systems}}, 
  year={2018},
  volume={17},
  number={6},
  pages={3834-3849},
  doi={10.1109/TWC.2018.2816929}}

@Article{Zeina2018,
  author  = {Mheich, Z. and Wen, Lei and Xiao, Pei and Maaref Amin},
  journal = {IEEE Trans. Veh. Technol.},
  title   = {{Design of SCMA Codebooks based on Golden Angle Modulation}},
  year    = {2018},
  month   = {Feb.},
  volumn  = {68},
  number  = {2},
  pages   = {1501–1509}
}

@Article{Yu2018,
  author  = {Yu, Lisu and Fan, Pingzhi and Cai, Dong and Ma, Zheng},
  journal = {IEEE Trans. Veh. Technol.},
  title   = {{Design and Analysis of SCMA Codebook based on Star-QAM Signaling Constellations}},
  year    = {2018},
  month   = {Nov.},
  volumn  = {67},
  number  = {11},
  pages   = {10543–10553}
}

@Article{Chen2020b,
  author  = {Chen, Yanming and Chen, JW},
  journal = {IEEE Trans. Commun.},
  title   = {{On the Design of Near-Optimal Sparse Code Multiple Access Codebooks}},
  year    = {2020},
  month   = {May},
  volumn  = {68},
  number  = {5},
  pages   = {2950–2962}
}

@Article{Liu2021,
  author  = {Liu, Zilong and Yang, Lie-Liang},
  journal = {IEEE Trans. Commun.},
  title   = {{Sparse or Dense: A Comparative Study of Code-Domain NOMA Systems}},
  year    = {2021},
  month   = {Aug},
  volumn  = {20},
  number  = {8},
  pages   = {4768-4780}
}

@Article{Mao2022,
  author  = {Mao, Yijie and Dizdar, Onur and Clerckx, Bruno and Schober, Robert and Popovski, Petar and Poor, H. Vincent},
  journal = {IEEE Commun. Surv. \& Tut.},
  title   = {{Rate-Splitting Multiple Access: Fundamentals, Survey, and Future Research Trends}},
  year    = {2022},
  number  = {4},
  pages   = {2073-2126},
  volume  = {24},
  doi     = {10.1109/COMST.2022.3191937},
}

@Article{Li2022,
  author   = {Li, Xudong and Gao, Zhicheng and Gui, Yiming and Liu, Zilong and Xiao, Pei and Yu, Lisu},
  journal  = {IEEE Trans. on Veh. Technol.},
  title    = {{Design of Power-Imbalanced {SCMA} Codebook}},
  year     = {2022},
  issn     = {1939-9359},
  month    = {Feb},
  number   = {2},
  pages    = {2140-2145},
  volume   = {71},
  doi      = {10.1109/TVT.2021.3132698},
}

@InProceedings{Deka_DE,
  author    = {Deka, Kuntal and Priyadarsini, Minerva and Sharma, Sanjeev and Beferull-Lozano, Baltasar},
  booktitle = {2020 IEEE Int. Conf. on Commun. Workshops (ICC Workshops)},
  title     = {{Design of {SCMA} Codebooks using Differential Evolution}},
  year      = {2020},
  month     = {June},
  pages     = {1-7},
  doi       = {10.1109/ICCWorkshops49005.2020.9145202},
  issn      = {2474-9133},
}

@InProceedings{Zhang2016,
  author    = {Zhang, Shutian and Xiao, Kexin and Xiao, Baicen and Chen, Zhiyong and Xia, Bin and Chen, Dageng and Ma, Shaodan},
  booktitle = {2016 8th Int. Conf. on Wireless Commun. \& Signal Process. (WCSP)},
  title     = {{A Capacity-based Codebook Design Method for Sparse Code Multiple Access Systems}},
  year      = {2016},
  month     = {Oct},
  pages     = {1-5},
  doi       = {10.1109/WCSP.2016.7752620},
  issn      = {2472-7628},
}

@InProceedings{Dizdar2020,
  author    = {Dizdar, Onur and Mao, Yijie and Han, Wei and Clerckx, Bruno},
  booktitle = {2020 IEEE 31st Annu. Int. Symp. on Pers., Indoor and Mobile Radio Commun.},
  title     = {{Rate-Splitting Multiple Access for Downlink Multi-Antenna Communications: Physical Layer Design and Link-level Simulations}},
  year      = {2020},
  pages     = {1-6},
  doi       = {10.1109/PIMRC48278.2020.9217326},
}

@INPROCEEDINGS{sic_free_conf,
  author={Zhang, Sibo and Clerckx, Bruno and Vargas, David},
  booktitle={2024 IEEE 25th Int. Workshop on Signal Process. Adv. in Wireless Commun. (SPAWC)}, 
  title={{SIC-Free Rate-Splitting Multiple Access: Constellation Constrained Sum-Rate Optimization}}, 
  year={2024},
  volume={},
  number={},
  pages={903-910},
  doi={10.1109/SPAWC60668.2024.10694077}}

@ARTICLE{rsma_primer,
  author={Clerckx, Bruno and Mao, Yijie and Jorswieck, Eduard A. and Yuan, Jinhong and Love, David J. and Erkip, Elza and Niyato, Dusit},
  journal={IEEE J. on Selected Areas in Commun.}, 
  title={{A Primer on Rate-Splitting Multiple Access: Tutorial, Myths, and Frequently Asked Questions}}, 
  year={2023},
  volume={41},
  number={5},
  pages={1265-1308},
  doi={10.1109/JSAC.2023.3242718}}

@ARTICLE{6g_survey,
  author={Wang, Cheng~Xiang and You, Xiaohu and Gao, Xiqi and Zhu, Xiuming and Li, Zixin and Zhang, Chuan and Wang, Haiming and Huang, Yongming and Chen, Yunfei and Haas, Harald and Thompson, John S. and Larsson, Erik G. and Renzo, Marco Di and Tong, Wen and Zhu, Peiying and Shen, Xuemin and Poor, H. Vincent and Hanzo, Lajos},
  journal={IEEE Commun. Surv. \& Tut.}, 
  title={{On the Road to {6G}: Visions, Requirements, Key Technologies, and Testbeds}}, 
  year={2023},
  volume={25},
  number={2},
  pages={905-974},
  doi={10.1109/COMST.2023.3249835}}

@ARTICLE{6g_tech,
  author={Zhang, Zhengquan and Xiao, Yue and Ma, Zheng and Xiao, Ming and Ding, Zhiguo and Lei, Xianfu and Karagiannidis, George K. and Fan, Pingzhi},
  journal={IEEE Veh. Technol. Mag.}, 
  title={{6G Wireless Networks: Vision, Requirements, Architecture, and Key Technology}}, 
  year={2019},
  volume={14},
  number={3},
  pages={28-41},
  doi={10.1109/MVT.2019.2921208}}

@ARTICLE{rs_2001,
  author={Grant, A.J. and Rimoldi, B. and Urbanke, R.L. and Whiting, P.A.},
  journal={IEEE Trans. on Inf. Theory}, 
  title={{Rate-Splitting Multiple Access for Discrete Memoryless Channels}}, 
  year={2001},
  volume={47},
  number={3},
  pages={873-890},
  doi={10.1109/18.915637}}

@INPROCEEDINGS{SCMA1,
  author={Nikopour, Hosein and Baligh, Hadi},
  booktitle={2013 IEEE 24th Annu. Int. Symp. on Pers., Indoor, and Mobile Radio Commun. (PIMRC)}, 
  title={{Sparse Code Multiple Access}}, 
  year={2013},
  volume={},
  number={},
  pages={332-336},
  doi={10.1109/PIMRC.2013.6666156}}

@INPROCEEDINGS{scma_cb,
  author={Taherzadeh, Mahmoud and Nikopour, Hosein and Bayesteh, Alireza and Baligh, Hadi},
  booktitle={2014 IEEE 80th Veh. Technol. Conf. (VTC2014-Fall)}, 
  title={{SCMA Codebook Design}}, 
  year={2014},
  volume={},
  number={},
  pages={1-5},
  doi={10.1109/VTCFall.2014.6966170}}

@ARTICLE{ldpc_5g,
  author={Richardson, Tom and Kudekar, Shrinivas},
  journal={IEEE Commun. Mag.}, 
  title={{Design of Low-Density Parity Check Codes for {5G} New Radio}}, 
  year={2018},
  volume={56},
  number={3},
  pages={28-34},
  doi={10.1109/MCOM.2018.1700839}}

@ARTICLE{RS_Gaussian,
  author={Rimoldi, B. and Urbanke, R.},
  journal={IEEE Trans. on Inf. Theory}, 
  title={{A Rate-Splitting Approach to the {G}aussian Multiple-Access Channel}}, 
  year={1996},
  volume={42},
  number={2},
  pages={364-375},
  doi={10.1109/18.485709}}

@ARTICLE{rsma_receiever,
  author={Zhang, Sibo and Clerckx, Bruno and Vargas, David and Haffenden, Oliver and Murphy, Andrew},
  journal={IEEE Trans. on Commun.}, 
  title={{Rate-Splitting Multiple Access: Finite Constellations, Receiver Design, and {SIC}-Free Implementation}}, 
  year={2024},
  volume={72},
  number={9},
  pages={5319-5333},
  doi={10.1109/TCOMM.2024.3383102}}

@ARTICLE{scma_zilong,
  author={Chaturvedi, Saumya and Liu, Zilong and Bohara, Vivek Ashok and Srivastava, Anand and Xiao, Pei},
  journal={IEEE Access}, 
  title={{A Tutorial on Decoding Techniques of Sparse Code Multiple Access}}, 
  year={2022},
  volume={10},
  number={},
  pages={58503-58524},
  doi={10.1109/ACCESS.2022.3178127}}

@ARTICLE{SCMA_det_low_complx,
  author={Yang, Lin and Liu, Yunyun and Siu, Yunming},
  journal={IEEE Commun. Lett.}, 
  title={{Low Complexity Message Passing Algorithm for SCMA System}}, 
  year={2016},
  volume={20},
  number={12},
  pages={2466-2469},
  doi={10.1109/LCOMM.2016.2609382}}

@INPROCEEDINGS{SCMA_det_low_complx1,
  author={Bayesteh, Alireza and Nikopour, Hosein and Taherzadeh, Mahmoud and Baligh, Hadi and Ma, Jianglei},
  booktitle={2015 IEEE Globecom Workshops (GC Wkshps)}, 
  title={{Low Complexity Techniques for SCMA Detection}}, 
  year={2015},
  volume={},
  number={},
  pages={1-6},
  doi={10.1109/GLOCOMW.2015.7414184}}

@ARTICLE{ioT_scma,
  author={Luo, Qu and Liu, Zilong and Chen, Gaojie and Xiao, Pei and Ma, Yi and Maaref, Amine},
  journal={IEEE Trans. on Wireless Commun.}, 
  title={{A Design of Low-Projection SCMA Codebooks for Ultra-Low Decoding Complexity in Downlink IoT Networks}}, 
  year={2023},
  volume={22},
  number={10},
  pages={6608-6623},
  doi={10.1109/TWC.2023.3244868}}

@INPROCEEDINGS{soft_SIC,
  author={Won-Joon Choi and Kok-Wui Cheong and Cioffi, J.M.},
  booktitle={2000 IEEE Wireless Commun. and Netw. Conf. Conf. Rec. (Cat. No.00TH8540)}, 
  title={{Iterative Soft Interference Cancellation for Multiple Antenna Systems}}, 
  year={2000},
  volume={1},
  number={},
  pages={304-309 vol.1},
  doi={10.1109/WCNC.2000.904647}}

@Inbook{Sharma2021,
author="Sharma, Sanjeev
and Deka, Kuntal",
editor="Mandloi, Manish
and Gurjar, Devendra
and Pattanayak, Prabina
and Nguyen, Ha",
title="Sparse Code and Hybrid Multiple Access Techniques",
bookTitle="5G and Beyond Wireless Systems: PHY Layer Perspective",
year="2021",
publisher="Springer Singapore",
address="Singapore",
pages="85--105",
isbn="978-981-15-6390-4",
doi="10.1007/978-981-15-6390-4_5",
url="https://doi.org/10.1007/978-981-15-6390-4_5"
}

@ARTICLE{zeng_low,
  author={Zeng, Weiliang and Xiao, Chengshan and Lu, Jianhua},
  journal={IEEE Wireless Commun. Lett.}, 
  title={{A Low-Complexity Design of Linear Precoding for MIMO Channels with Finite-Alphabet Inputs}}, 
  year={2012},
  volume={1},
  number={1},
  pages={38-41},
  doi={10.1109/WCL.2012.121411.110025}}

@ARTICLE{Chen2020_mc_rsma,
  author={Chen, Hongzhi and Mi, De and Clerckx, Bruno and Chu, Zheng and Shi, Jia and Xiao, Pei},
  journal={IEEE Trans. on Vehicular Technology}, 
  title={{Joint Power and Subcarrier Allocation Optimization for Multigroup Multicast Systems With Rate Splitting}}, 
  year={2020},
  volume={69},
  number={2},
  pages={2306-2310},
  doi={10.1109/TVT.2019.2957994}}

@ARTICLE{6g_new2,
  author={Bariah, Lina and Mohjazi, Lina and Muhaidat, Sami and Sofotasios, Paschalis C. and Kurt, Gunes Karabulut and Yanikomeroglu, Halim and Dobre, Octavia A.},
  journal={IEEE Access}, 
  title={{A Prospective Look: Key Enabling Technologies, Applications and Open Research Topics in 6G Networks}}, 
  year={2020},
  volume={8},
  number={},
  pages={174792-174820},
  doi={10.1109/ACCESS.2020.3019590}}

@ARTICLE{6g_acces,
  author={Bhat, Jagadeesha R. and Alqahtani, Salman A.},
  journal={IEEE Access}, 
  title={{6G Ecosystem: Current Status and Future Perspective}}, 
  year={2021},
  volume={9},
  number={},
  pages={43134-43167},
  doi={10.1109/ACCESS.2021.3054833}}

@ARTICLE{BP,
  author={Kschischang, F.R. and Frey, B.J. and Loeliger, H.-A.},
  journal={IEEE Trans. Inf. Theory}, 
  title={{Factor graphs and the sum-product algorithm}}, 
  year={2001},
  volume={47},
  number={2},
  pages={498-519},
  doi={10.1109/18.910572}}
\end{document}